\documentclass{PoS}

\usepackage{graphicx}
\usepackage{epstopdf}
\usepackage{amsmath,amssymb}
\usepackage{cancel}

\newcommand{\beq}{\begin{equation}}
\newcommand{\eeq}{\end{equation}}
\newcommand{\bea}{\begin{eqnarray}}
\newcommand{\eea}{\end{eqnarray}}
\newcommand{\nn}{\nonumber}

\newcommand{\fig}{Fig.~}

\newcommand{\bx}{{\bf x}}
\newcommand{\by}{{\bf y}}

\def\lsi{\raise0.3ex\hbox{$<$\kern-0.75em\raise-1.1ex\hbox{$\sim$}}}
\def\gsi{\raise0.3ex\hbox{$>$\kern-0.75em\raise-1.1ex\hbox{$\sim$}}}
\newcommand{\lsim}{\mathop{\lsi}}
\newcommand{\gsim}{\mathop{\gsi}}

\title{Constraining the QCD phase diagram at finite temperature and density}

\ShortTitle{The QCD phase diagram}

\author{\speaker{Owe Philipsen}\\
        Institut für Theoretische Physik, Goethe-Universität Frankfurt am Main\\
        Max-von-Laue-Str. 1, 60438 Frankfurt am Main, Germany\\
        E-mail: \email{philipsen@th.physik.uni-frankfurt.de} }        
        
\abstract{
Neither the chiral limit nor finite baryon density can be simulated directly in lattice QCD,
which severely limits our understanding of the QCD phase diagram.
In this review I collect results for the phase structure in an extended parameter space 
of QCD, with varying numbers of flavours, quark masses, colours, lattice spacings, imaginary and
isospin chemical potentials. Such studies help in understanding the underlying symmetries 
and degrees of freedom, and are beginning to provide a consistent picture constraining the possibilities for the physical
phase diagram.
}

\FullConference{37th International Symposium on Lattice Field Theory - Lattice2019\\
		16-22 June 2019\\
		Wuhan, China}

\begin{document}

\section{Introduction}

An outstanding open problem of QCD is the nature of its phase diagram as a function of temperature and
baryon chemical potential. A change in dynamics from a hadron gas to a regime governed by
different degrees of freedom, as in a quark gluon plasma, is expected to be caused by an effective restoration of chiral
symmetry in a region with temperatures $T\lsim 160$ MeV and baryon chemical potential $\mu_B\lsim 1$ GeV. Such low energy scales
necessitate a non-perturbative first-principles approach like lattice QCD. Unfortunately, a severe sign problem prohibits
simulations by importance sampling for non-vanishing chemical potential \cite{deForcrand:2010ys}. 
Despite tremendous efforts over several decades, no genuine solution to this problem is available.  

In this contribution I will not cover the sign problem and the many attempts to alleviate it
algorithmically. Instead I will report on a different strategy to learn about the phase diagram, which is to consider QCD thermodynamics in 
the parameter space $\{T,\mu_B,\mu_I, N_f,m_q^f,g^2,N_c\}$.
By studying the phase structure in every parameter region, where some or another method works,
an increasing number of constraints on the physical QCD phase diagram is obtained. As a by-product, such studies
also provide theoretical insight about the interplay of the involved symmetries and degrees of freedom. 
I will begin with a discussion of the nature of the chiral phase transition in the chiral limit, before turning to proper finite density.

\section{The Columbia plot and its extended versions}

\subsection{The order of the thermal transition at zero density \label{sec:columbia}}

\begin{figure}[t]
\vspace*{-0.75cm}
\centering
\includegraphics[width=\textwidth]{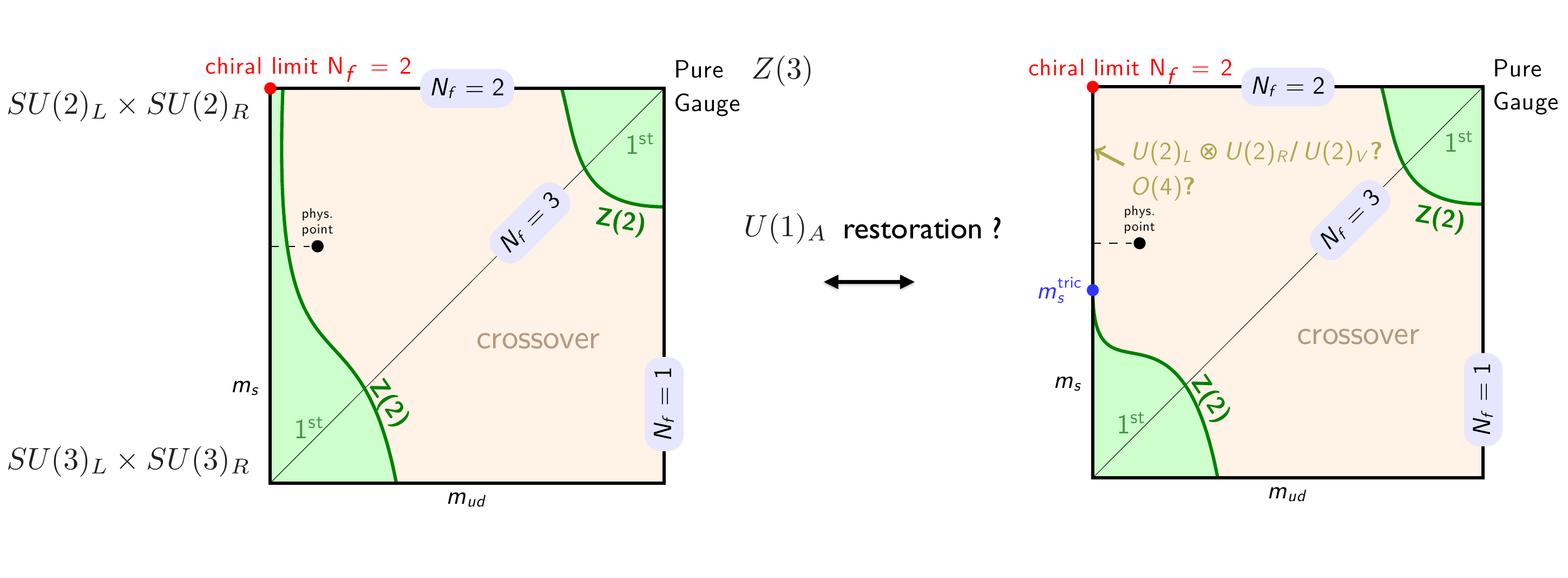}
\vspace*{-1.25cm}
\caption[]{Possible scenarios for the order of the thermal QCD transition as a function of the quark masses.}
\label{fig:columbia}
\end{figure}
The nature of the thermal QCD transition with $N_f=2+1$ quark flavours as a function of the quark masses is
summarised in the so-called Columbia plot, \fig\ref{fig:columbia}. In the quenched limit QCD reduces to $SU(3)$ Yang-Mills theory in
the presence of static quarks and shows a first-order phase transition \cite{boyd}
associated with the spontaneous breaking of the $Z(3)$ center
symmetry. Once quarks have a finite mass, the center symmetry is explicitly broken and the first-order phase transition weakens
as the quarks get lighter, until it is lost at a $Z(2)$ second-order line of critical mass values. 

In the opposite, chiral limit, the situation is more complicated, and for a long time expectations have been mostly guided by 
an analysis based on the epsilon expansion \cite{piwi}. It predicts the chiral
transition to be first-order for $N_f\geq 3$, whereas the case of $N_f=2$ is found to crucially depend
on the fate of the anomalous $U(1)_A$ symmetry: if the latter remains broken at $T_c$, the chiral transition should be second
order in the $O(4)$-universality class, whereas its effective restoration would enlarge the chiral symmetry and push the transition to first-order.
A later high order perturbative analysis of renormalisation group flow \cite{Pelissetto:2013hqa} instead finds a possible 
symmetry breaking pattern
to be $U(2)_L\otimes U(2)_R\rightarrow U(2)_V$ in the case of a restored $U(1)_A$, 
which would amount to a second-order transition in a different
universality class. For non-zero quark masses, chiral symmetry is explicitly broken and a first-order chiral transition weakens to disappear
at a $Z(2)$ second-order critical boundary, while a second-order transition disappears immediately.

Computing these boundaries numerically is punishingly expensive. Locating a transition requires scans in temperature and mass,
deciding its order and universality requires finite size scaling analyses with sufficiently large and different volumes.  There is 
critical slowing down near a critical point as well as approaching the continuum, and the required quark masses are mostly smaller
than physical.
On coarse $N_\tau=4$ lattices, the first-order region is explicitly seen for $N_f=3$ unimproved 
staggered \cite{Karsch:2001nf,deForcrand:2003vyj} as well as $O(a)$-improved Wilson \cite{Jin:2014hea} fermions, the
narrower $N_f=2$ region with unimproved staggered \cite{Bonati:2014kpa,Cuteri:2017gci} and 
unimproved Wilson \cite{Philipsen:2016hkv} fermions.
However, the location of the boundary
line varies widely between these, indicating large cut-off effects. On the other hand, simulations with an improved staggered 
action (HISQ) do not see a first-order region on $N_\tau=6$ lattices even for $N_f=3$ \cite{Bazavov:2017xul}.
The only point with a continuum extrapolation (besides the pure gauge limit) is the physical point, which has been identified to 
be in the crossover region \cite{aoki06}.

\subsection{The chiral transition as a function of $N_f$}

\begin{figure}[t]
\centering
\vspace*{-1.0cm}
\includegraphics[width=0.4\textwidth]{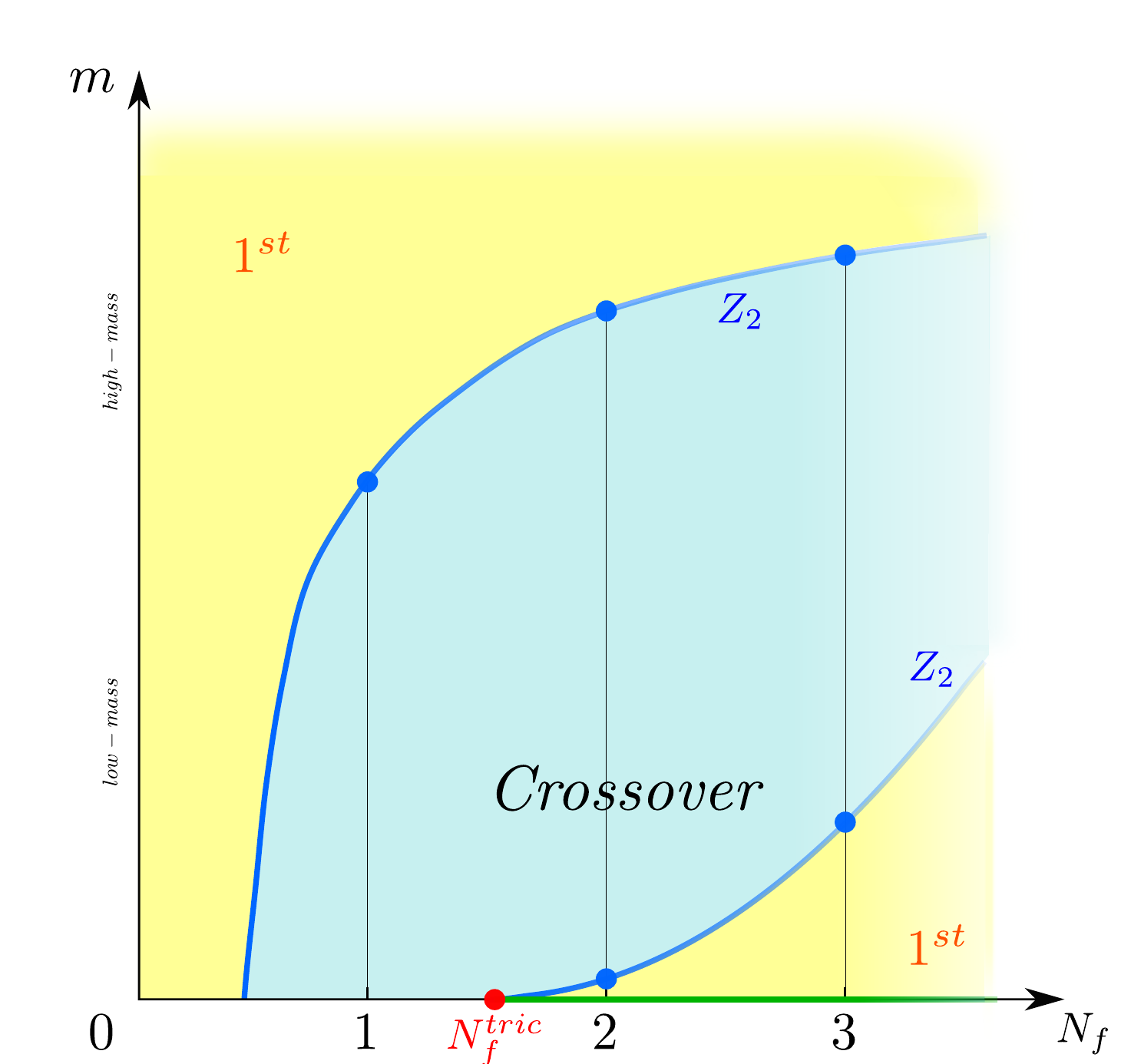}
\includegraphics[width=0.5\textwidth]{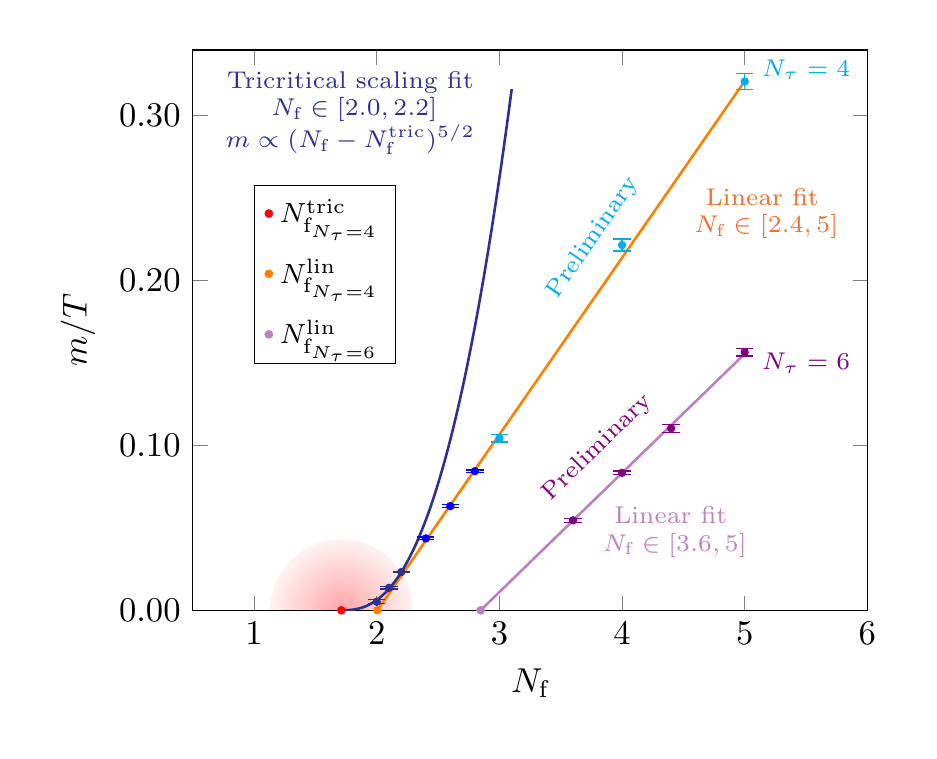}
\caption[]{Left: Order of the thermal transition as function of $N_f$ and quark mass (schematic). 
Right: Second-order chiral critical line
from unimproved staggered fermions \cite{Cuteri:2018wci}.}
\label{fig:nf_columbia}
\end{figure}
In an alternative version of the Columbia plot, the chiral transition is considered as a function of 
the number of degenerate quark flavours and their mass. The first scenario from \fig\ref{fig:columbia} then translates into
\fig\ref{fig:nf_columbia} (left). The chiral symmetry and, with it, the strength of the transition increases with the number of flavours.
One may now consider a partition function with the quark determinant raised to 
continuous, non-integer powers
of $N_f$, in order to study the approach to the chiral limit  \cite{Cuteri:2017gci}. For sufficiently large $N_f$, 
the chiral limit corresponds to a first-order, three-state coexistence line
(with $\langle \bar{\psi}\psi\rangle=0,\pm \mathrm{const}$). The weakening of this transition with decreasing $N_f$ implies 
a tricritical point, in which it ends. The chiral critical line, which separates 
the first-order from the crossover region, enters the tricritical point with a known tricritical exponent. Of course, there is no physical meaning in this non-integer 
value of $N_f^\mathrm{tric}$ other than to be smaller or larger than two, which puts the chiral transition for $N_f=2$ in the first- or
second-order region, respectively. 

A study on $N_\tau=4$ with unimproved staggered fermions \cite{Cuteri:2017gci} fully confirms these 
considerations, cf.~\fig\ref{fig:nf_columbia} (right). For sufficiently small masses tricritical scaling is observed and 
can be used to extrapolate to the chiral limit. Unfortunately, the scaling region is small and little is gained over 
direct simulations at $N_f=2$.  On $N_\tau=6$ lattices 
the scaling region is too low in quark mass to be simulated straightforwardly. 
However, beyond the scaling region the chiral critical line is observed to be linear 
over a large range of $N_f$-values. A linear extrapolation will then produce an upper bound for $N_f^\mathrm{tric}$. 
In order to draw conclusions, one would need to know the size of the scaling region, where the chiral critical line is curved.
In any case, a strong trend is seen for all $N_f$: decreasing the lattice spacing dramatically shrinks the chiral critical
quark mass bounding a first-order chiral transition.  
Similar observations up to $N_\tau=10$ are made for $N_f=4$ \cite{deForcrand:2017cgb}
and also for $N_f=3$ with $O(a)$-improved Wilson quarks \cite{Jin:2017jjp} \fig\ref{fig:2+1_scaling} (left). 

While these calculations do not yet permit unambiguous continuum extrapolations,
they provide a consistent picture across all discretisations: 
the chiral transition probed by lattice simulations weakens considerably as the
continuum is approached. This makes a second-order scenario for the $N_f=2$ chiral limit more likely and raises
the interesting question, whether also $N_f=3,4$ might possibly have second-order transitions, contrary to what was expected
for a long time. 

\subsection{From the physical point towards the chiral limit \label{sec:scale}}

\begin{figure}[t]
\centering
\vspace*{-0.5cm}
\includegraphics[width=0.5\textwidth]{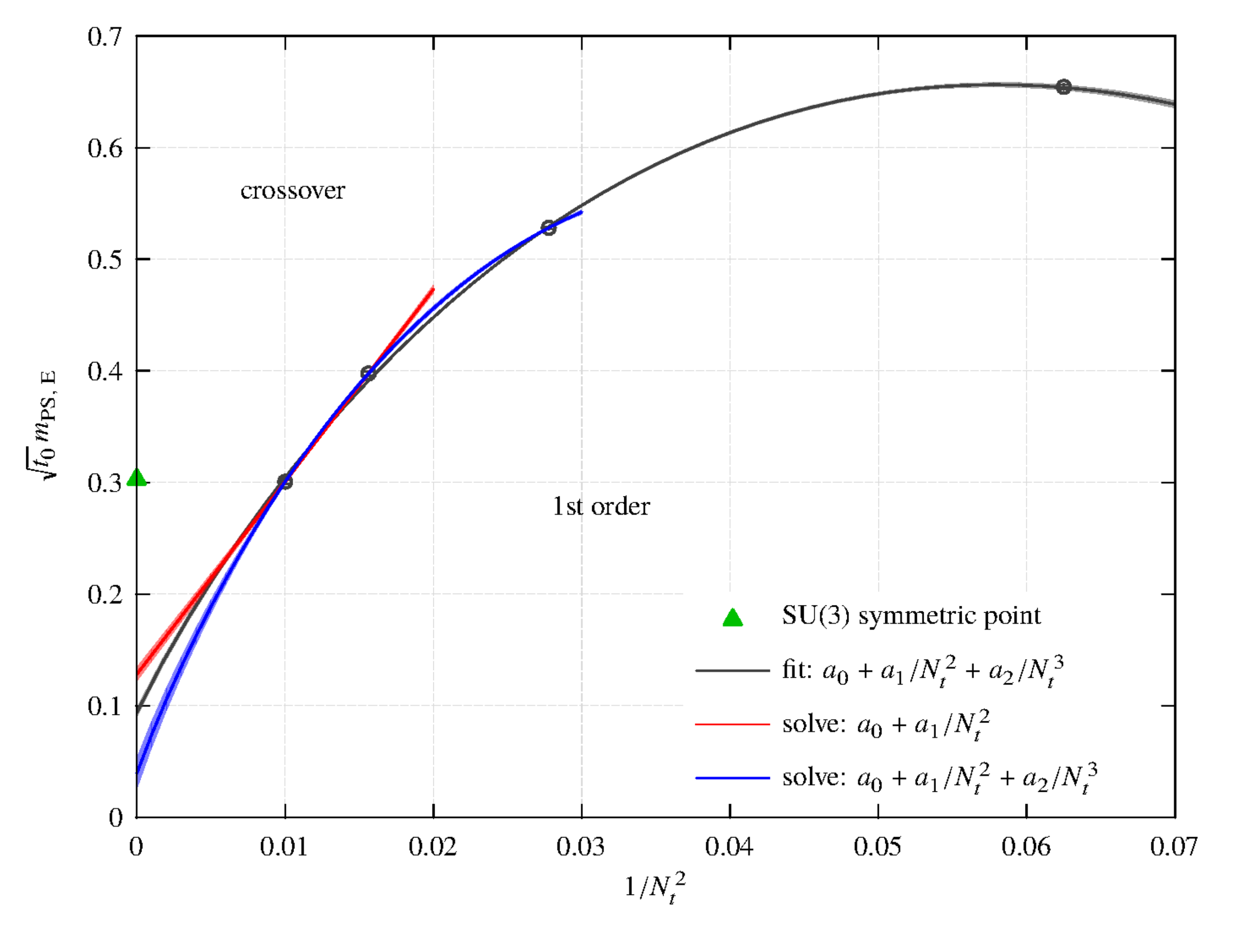}\hspace*{1cm}
\includegraphics[width=0.45\textwidth]{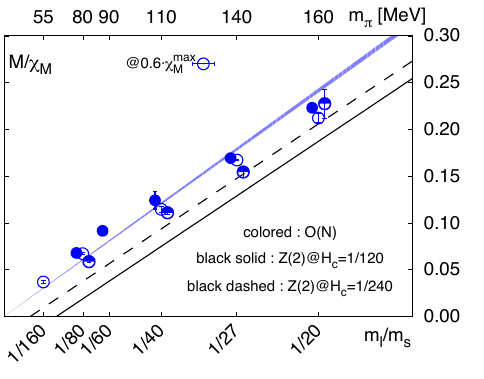}
\caption[]{Left: Chiral critical point for $N_f=3$ $O(a)$ improved Wilson fermions \cite{Jin:2017jjp}.
Right: Magnetic equation of state approaching the chiral limit. Lines represent fits to $O(N)$- (second-order scenario) or
$Z(2)$-scaling with a finite critical quark mass (first-order scenario) \cite{Ding:2019fzc}. 
}
\label{fig:2+1_scaling}
\end{figure}
There is then no contradiction between unimproved and improved staggered actions, which 
do not see a first-order region so far. Recent investigations searching for the chiral phase transition with improved actions 
start at the $N_f=2+1$ physical point and then reduce the light quark masses. Thus either
a $Z(2)$-critical point bounding the first-order region is approached, or a second-order transition in the chiral limit.  

In \cite{Ding:2019fzc,Ding:2019prx} the scaling behaviour was checked 
as the chiral limit is approached. Simulations were
carried out using the HISQ action on lattices with $N_\tau=6,8,10$ and light quark masses down to $m_\pi\approx 55$ MeV.
The analysis uses a renormalisation group invariant combination of chiral condensates as order parameter representing the magnetisation,
and the light quark masses in units of the strange quark mass as symmetry breaking field,
\beq
M=2(m_s\langle\bar{\psi}\psi\rangle_l-m_l\langle \bar{\psi}\psi\rangle_s)/f_K^4,\quad H=m_{l}/m_s\;.
\label{eq:scaling}
\eeq 
Near a critical point the magnetic equation of state is then dominated by universal scaling functions 
\beq
M(t,h)=h^{1/\delta}f_G(z)+\ldots\;, \quad \chi_M(t,h)=\frac{\partial M}{\partial H}=h_0^{-1}h^{1/\delta-1}f_\chi(z)+\ldots\;,
\eeq
with a scaling variable $z=t/h^{1/\beta\delta}$ expressed in terms of the reduced temperature and external field, 
$t=t_0(T-T^0_c)/T^0_c, h=H/h_0$, which contain the unknown critical temperature in the chiral limit, $T_c^0$, and 
two non-universal parameters $t_0,h_0$.

\fig\ref{fig:2+1_scaling} (left) shows the ratio $M/\chi_M$,
whose approach to a critical point is sensitive to the scaling functions pertaining to the appropriate universality class. 
While it is difficult to distinguish between $O(4)$ and $Z(2)$, it is apparent that any finite critical quark mass bounding
a first-order region has to be excessively small to be consistent with the observed behaviour.
There is a similar scaling expression for the approach of the pseudo-critical crossover temperature of some observable $X$ to the critical
temperature in the chiral limit,
\beq
T_X(H)=T_c^0+\frac{z_X}{z_0}T_c^0H^{1/\beta\delta}\;, \quad 
T_c^0=132^{+3}_{-6} \mathrm{MeV}\;.
\label{eq:tc}
\eeq
In \cite{Ding:2019prx} the variation with the possible critical exponents is observed to be very small, so that an
extrapolation makes sense even without definite knowledge of the eventual universality class.
Results were checked to be
stable when the continuum extrapolation is done before the chiral extrapolation, leading to the critical temperature 
as a first result for the chiral limit.
Note that its value is $\sim 25$ MeV lower than the pseudo-critical temperature at the physical point, 
which should be important for phenomenological
descriptions of chiral symmetry breaking. 

Similar conclusions are drawn in an exploratory study attempting to approach the chiral 
limit more economically by reweighting in the light quark mass \cite{Endrodi:2018xto}, with a weight factor
\beq
W=\frac{\det(D)}{\det(D+m_{ud})}=\exp\Big[-\frac{V}{T}m_{ud}\bar{\psi}\psi(m_{ud})+O(m_{ud}^4)\Big] \;.
\eeq
The Banks-Casher relation is used to relate the chiral condensate
to the spectral density of the Dirac operator, which is then reweighted to zero mass. 
Note, that this involves an infinite volume extrapolation
by polynomial fits, whose systematics still needs studying. Nevertheless, 
the finite size scaling of the chiral condensate obtained in this approach clearly prefers a second-order scenario.

\subsection{The $U(1)_A$ anomaly}

As pointed out in section \ref{sec:columbia}, the fate of the anomalous $U(1)_A$ around the thermal transition is expected to play
a significant role in determining the order of the chiral phase transition. Simulations of a model realising the QCD chiral
symmetry with a tuneable strength of the $U(1)_A$ anomaly indeed demonstrate that a first-order transition occurs for restored symmetry,
which changes to an $O(4)$-transition above some critical strength of symmetry breaking \cite{Chandrasekharan:2007up}. 
Unfortunately, this "critical strength'' of symmetry breaking is nothing universal that could be easily mapped to QCD,
which thus has to be investigated directly.
\begin{figure}[t]
\centering
\vspace*{-0.5cm}
\includegraphics[height=4.5cm]{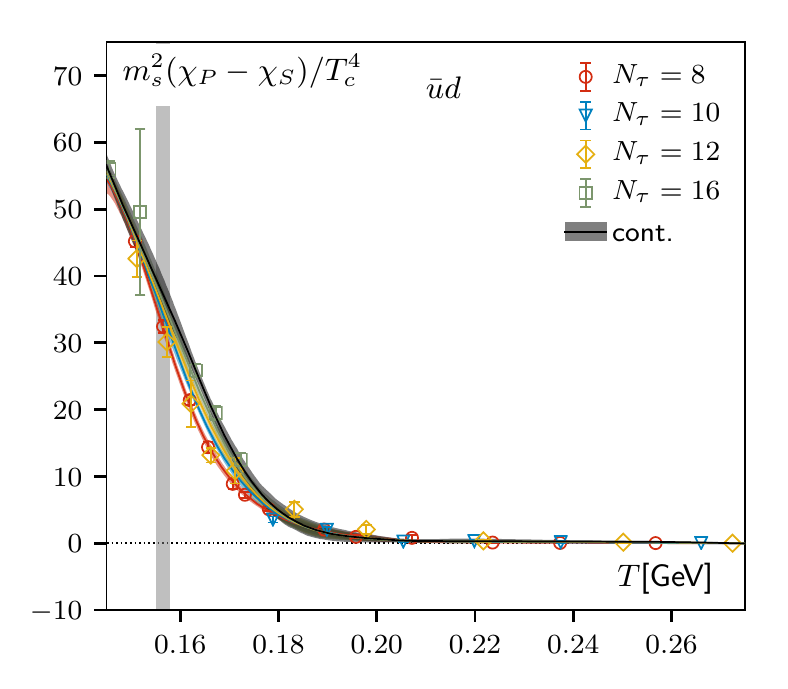}\hspace*{1.5cm}
\includegraphics[height=4.5cm]{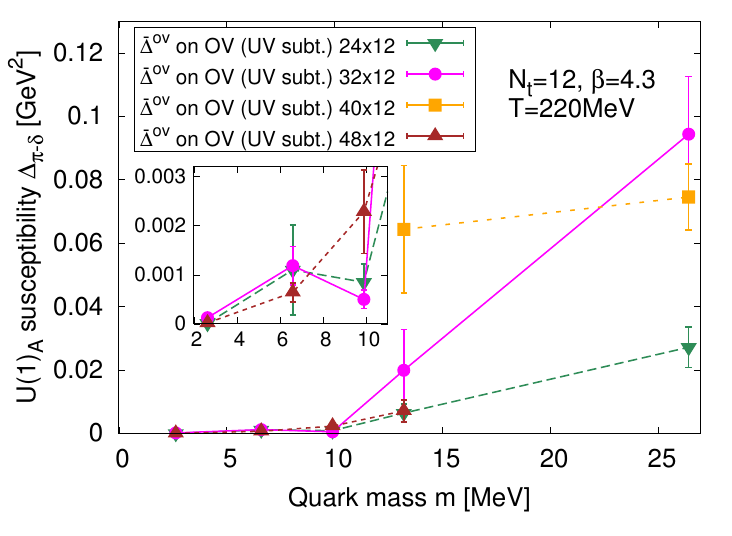}
\caption[]{Splitting between scalar and pseudo-scalar susceptibilities. Left: Overlap operator evaluated on HISQ ensembles at
the physical point \cite{hegde}. Right: DW fermions reweighted to overlap \cite{Suzuki:2017ifu}.}
\label{fig:u1a}
\end{figure}

Several attempts over the last years used using different discretisation schemes, 
from which apparently contradicting conclusions were drawn.  
Studies with chiral domain wall fermions on $N_\tau=8$ with $200$ MeV pions \cite{Buchoff:2013nra} find evidence of a broken $U(1)_A$ at 
the pseudo-critical temperature, and the same result is reported when evaluating the overlap operator on 
$N_\tau=8$ HISQ ensembles with $160$ MeV pions \cite{Dick:2015twa}. 
On the other hand, the screening mass spectrum evaluated at the pseudo-critical
temperature with $O(a)$-improved Wilson fermions on fine $N_\tau=16$ lattices suggests a significant reduction of the 
anomaly in the chiral limit \cite{Brandt:2016daq}, and M\"obius domain wall fermions reweighted to overlap on $N_\tau=8,10$ 
show full symmetry restoration
at $T\sim 220$ MeV.  Updates on these results, using the difference of scalar and pseudo-scalar susceptibilities
as a measure for the anomaly, 
\beq
\Delta_{\pi-\delta}=\chi_\pi-\chi_\delta=\int d^4x\; \langle \pi^a(x)\pi^a(0) -\delta^a(x)-\delta^a(0)\rangle\;,
\eeq
are shown in \fig\ref{fig:u1a}. On the left the overlap Dirac operator is
evaluated on HISQ ensembles with physical quark masses, including an extrapolation to the continuum limit \cite{hegde}. 
The full restoration of the symmetry happens above the chiral crossover temperature, but this is not yet
the chiral limit. \fig\ref{fig:u1a} (right) shows
the same quantity as a function quark mass, with domain wall fermions reweighted to overlap on $N_\tau=12$ at 
$T=220$ MeV \cite{Suzuki:2017ifu}.
The splitting is observed to vanish in the chiral limit, but in this case the temperature is above the transition.
The qualitative features of both calculations are fully consistent and no contradiction is apparent yet. 
It would be most valuable to compare both approaches for a set of identical 
parameter values. For future investigations it might also be useful to check further
relations between chiral and $U(1)_A$ restoration, provided by certain Ward identities, which give information on the scaling with 
temperature as a criterion for exact symmetry restoration \cite{Nicola:2018itc}.

\subsection{QCD with imaginary chemical potential}

\begin{figure}[t]
\vspace*{-0.75cm}
\centering
\includegraphics[width=0.35\textwidth]{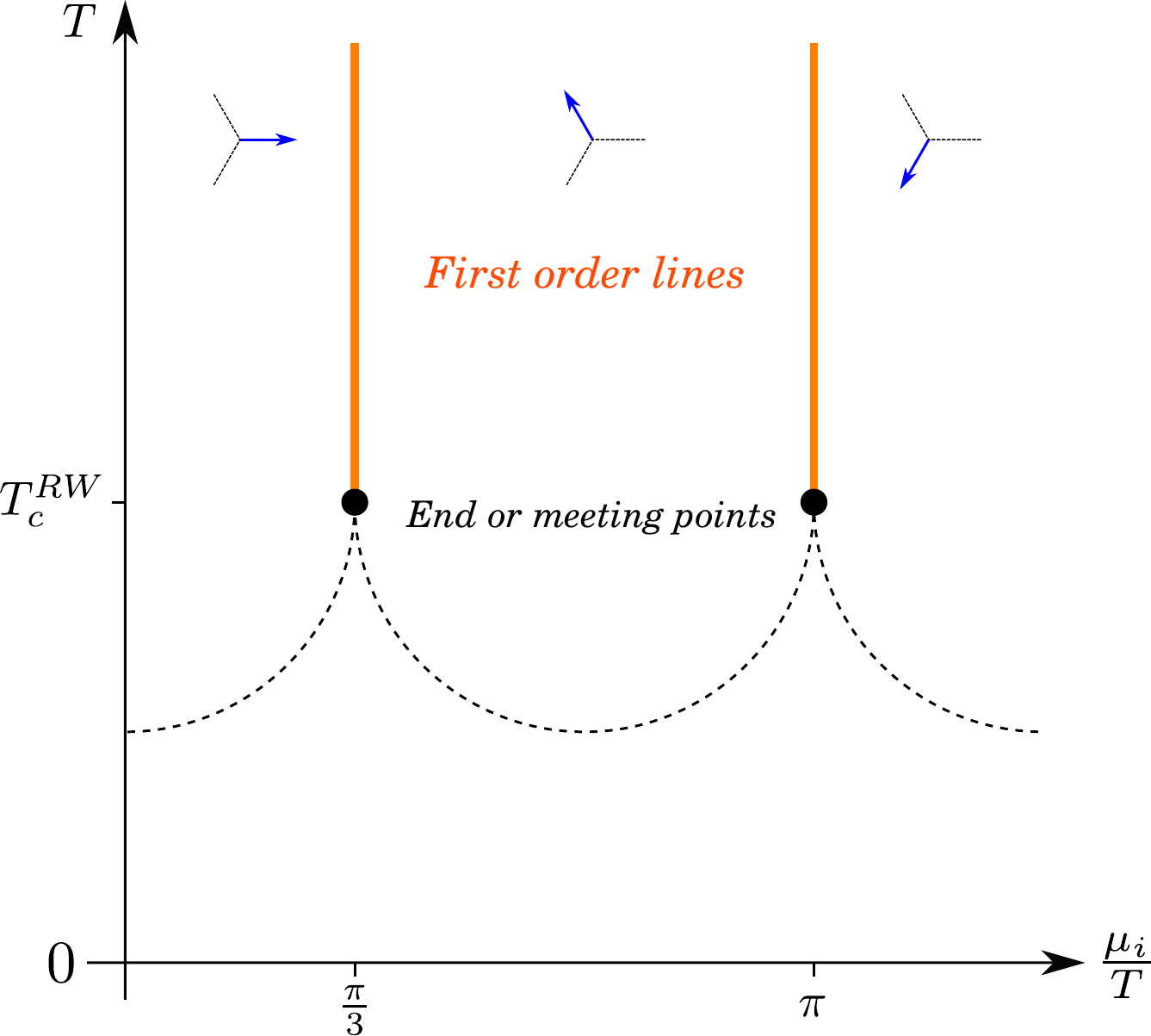}\hspace*{1.0cm}
\includegraphics[width=0.4\textwidth]{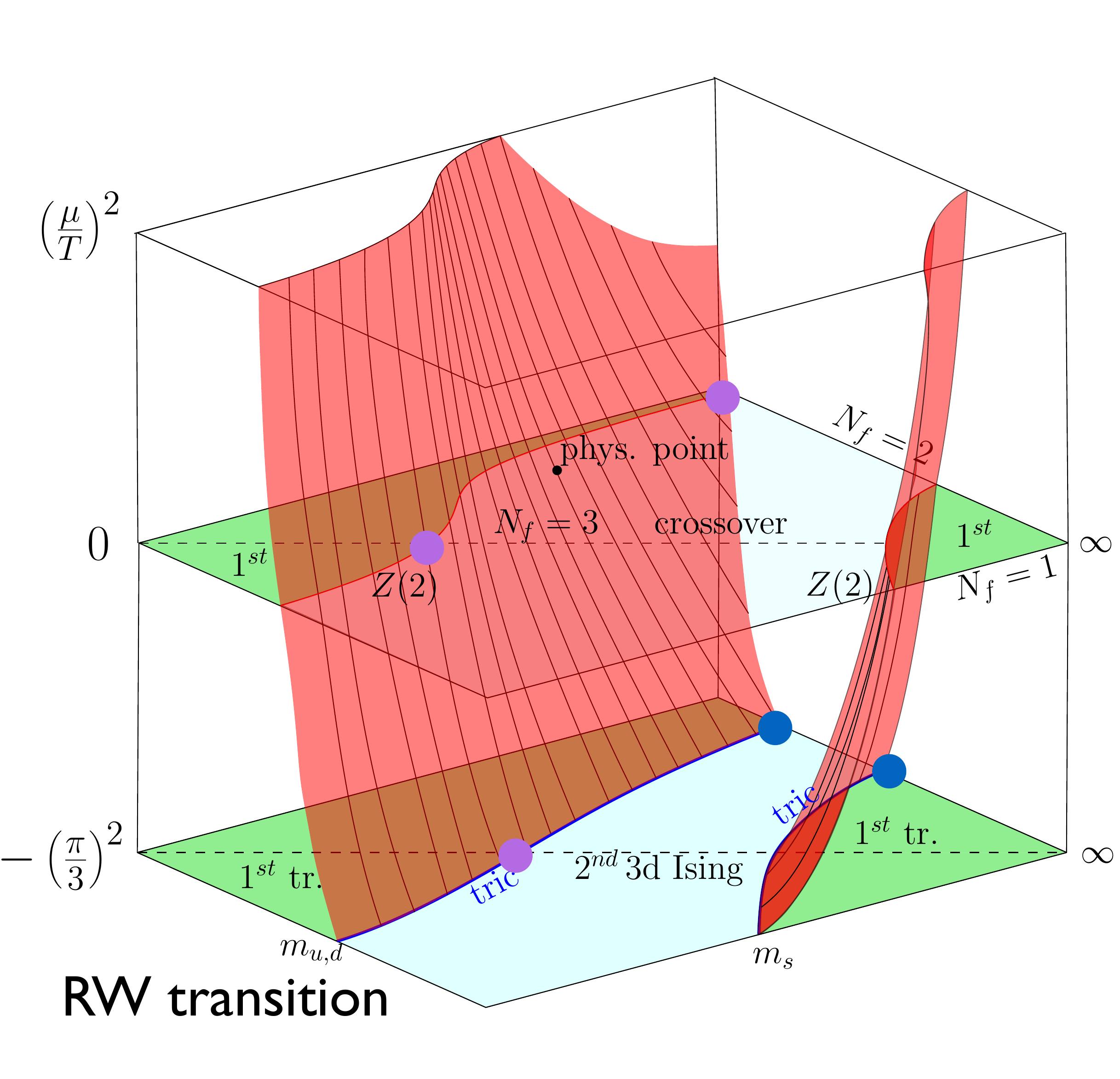}
\caption[]{Left: The QCD phase diagram at imaginary chemical potential. Vertical lines mark first-order transitions between different
center sectors, the dotted lines are the analytic continuation of the transitions at real $\mu$, whose nature depends on the quark masses. 
Right: Columbia plot with chemical potential. The bottom plane corresponds to the first center-transition.}
\label{fig:3dcolumbia}
\end{figure}
While it is unphysical, imaginary chemical potential for quark number, $\mu=i\mu_i, \mu_i\in \mathbb{R}$, does not induce a sign
problem and thus can be simulated without difficulty. This has been used to extract several aspects of the low density phase diagram 
at real $\mu$ by analytic continuation \cite{deForcrand:2002hgr,DElia:2002tig}. Two exact symmetries facilitate such studies. Because
of $CP$-invariance, the QCD partition function is even in chemical potential $Z(\mu)=Z(-\mu)$. Furthermore, for arbitrary fermion masses
it is periodic in imaginary chemical potential because of the global Roberge-Weiss (or center) symmetry \cite{Roberge:1986mm},
\beq
Z\left(T,i\frac{\mu_i}{T}\right)=Z\left(T,i\frac{\mu_i}{T}+i\frac{2\pi n}{N_c}\right)\;.
\eeq
This leads to transitions cycling through the $N_c$ center sectors, which are distinguishable by the phase of the Polyakov loop but have 
equal thermodynamic functions. The transitions are first order
for high temperatures and crossover for low temperatures, see \fig\ref{fig:3dcolumbia} (left).
In the $T$-direction there is the analytic continuation of the QCD thermal transition, whose order depends on $N_f$ and the quark masses. 
For first-order chiral and deconfinement transitions (small and large quark masses), the transition lines
meet up in a triple-point, while for thermal crossover the RW-transition ends in a critical endpoint with 3d Ising universality. The boundary
between these scenarios is marked by a tricritical point. These structures have been established explicitly 
for unimproved staggered \cite{deForcrand:2010he,Bonati:2010gi} as well as unimproved Wilson \cite{Philipsen:2014rpa} fermions.
 
This translates into a 3d extension of the Columbia plot, as in \fig\ref{fig:3dcolumbia} (right). Regions of chiral
 and deconfinement phase transitions are now separated
by critical surfaces from the crossover region. The curvature of the chiral critical surface has been shown to be negative
both on the $N_f=3$ diagonal, as well as near the physical point \cite{deForcrand:2006pv} on $N_\tau=4$ lattices. 
Thus the chiral transition strengthens with imaginary and weakens 
with real chemical potential. This is opposite to a scenario with a chiral critical point close to the temperature axis, which
would require the chiral transition at the physical point to stregthen with real $\mu$.
Unfortunately,
because of the receding first-order region, these calculations could not yet be repeated on finer lattices.
\begin{figure}[t]
\centering
\vspace*{-0.5cm}
\hspace*{-0.5cm}
\includegraphics[height=4cm]{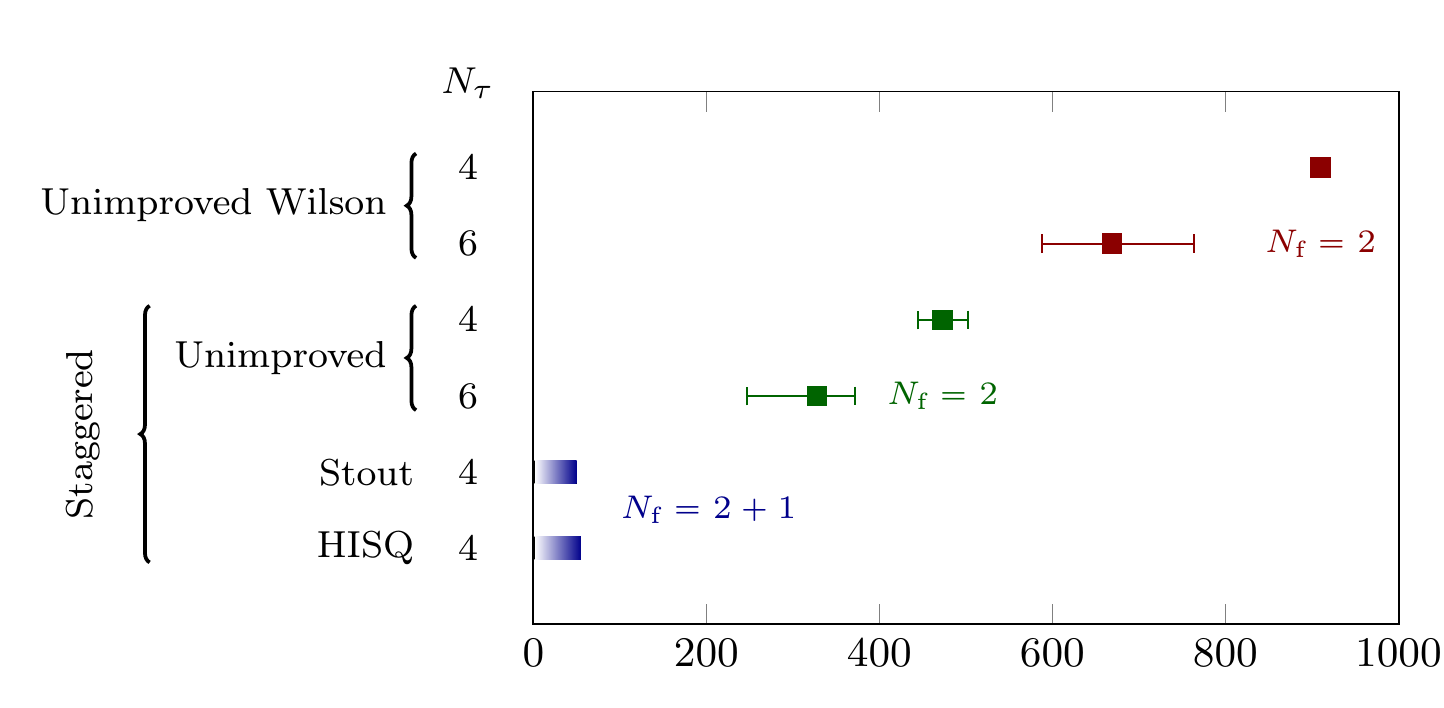}
\put(2,7){\tiny MeV}
\hspace*{0.5cm}
\includegraphics[height=4.5cm]{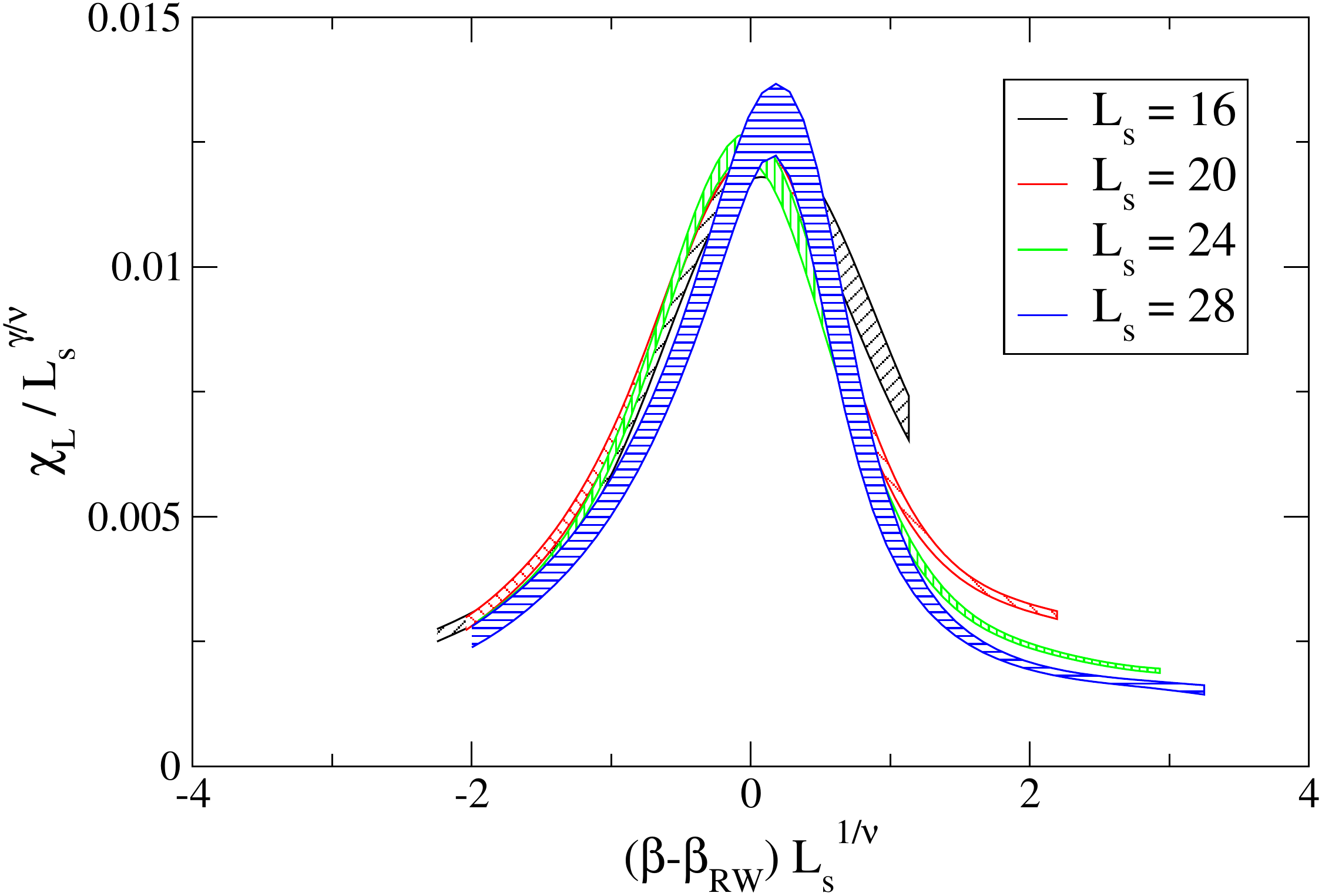}
\caption[]{Left: Tricritical pion mass values delimiting the first-order chiral region in the RW-plane \cite{Philipsen:2019ouy}. 
Right: Finite size scaling with 3d $Z(2)$ exponents for stout-smeared staggered fermions with 
$m_\pi\approx 50$ MeV and $(m_{ud}/m_s)_\mathrm{phys}$ on $N_\tau=4$ in the RW-plane \cite{Bonati:2018fvg}.}
\label{fig:rw_scale}
\end{figure}

Investigations on finer lattices reveal the same trend as seen at $\mu=0$, namely the chiral tricritical line moving towards
smaller quark masses, both for unimproved staggered \cite{Philipsen:2019ouy} and Wilson \cite{Cuteri:2015qkq} quarks, 
\fig\ref{fig:rw_scale} (left).
For stout-smeared staggered \cite{Bonati:2018fvg} and HISQ \cite{Goswami:2019exb} actions on $N_\tau=4$,
even the larger first-order region in the RW-plane cannot be detected when starting from the physical point and
reducing the pion masses down to $m_\pi\approx 50$ MeV, as \fig\ref{fig:rw_scale} (right) demonstrates with 
second-order scaling. 

Together with the $\mu=0$ results, this means that the entire chiral critical surface is shifting drastically towards the chiral limit as the lattice
spacing is decreased, and it is an open question whether any first-order transition remains in the continuum limit.
At the same time, this implies a softening of the crossover at the physical point, with so far no indication of a chiral 
critical structure at real chemical potential. 
 
\section{QCD at the physical point}

\subsection{Emerging chiral spin symmetry}

Besides the location and nature of phase transitions, studies of phase diagrams are also concerned with an
identification of the dominant dynamical degrees of freedom in each regime, which are expected to reflect the underlying symmetries.
In QCD, in particular, the hadronic regime is usually associated with broken chiral symmetry, while the quark gluon plasma represents
a symmetry restored state. In this context new investigations point to an interesting intermediate temperature regime with an emerging
$SU(4)$ symmetry, which was first proposed in \cite{Glozman:2014mka}.

Consider a $SU(2)_{CS}$ chiral spin transformation of quark fields defined by
\beq
\psi(x)\rightarrow \exp\Big( i\vec{\Sigma} \cdot\vec{\epsilon}\Big)\;,
\quad \Sigma_k=\{\gamma_k,-i\gamma_5\gamma_k,\gamma_5\}\;.
\eeq   
The QCD Lagrangian is not invariant under such transformations. However, when there is a thermodynamic medium implying a preferred
Lorentz frame, one finds the colour-electric part of the quark-gluon interaction as well as a chemical potential term for fermion number to 
be invariant, while kinetic terms and colour-magnetic interactions are not. 
Combining chiral spin symmetry with isospin, $SU(2)_{CS}\times SU(2)$, it
can be embedded in a $SU(4)$ symmtery that fully contains the usual chiral symmetry of the Lagrangian, 
$SU(4)\supset SU(2)_L\times SU(2)_R\times U(1)_A$.

\begin{figure}[t]
\vspace*{-0.5cm}
\centering
\includegraphics[width=0.32\textwidth]{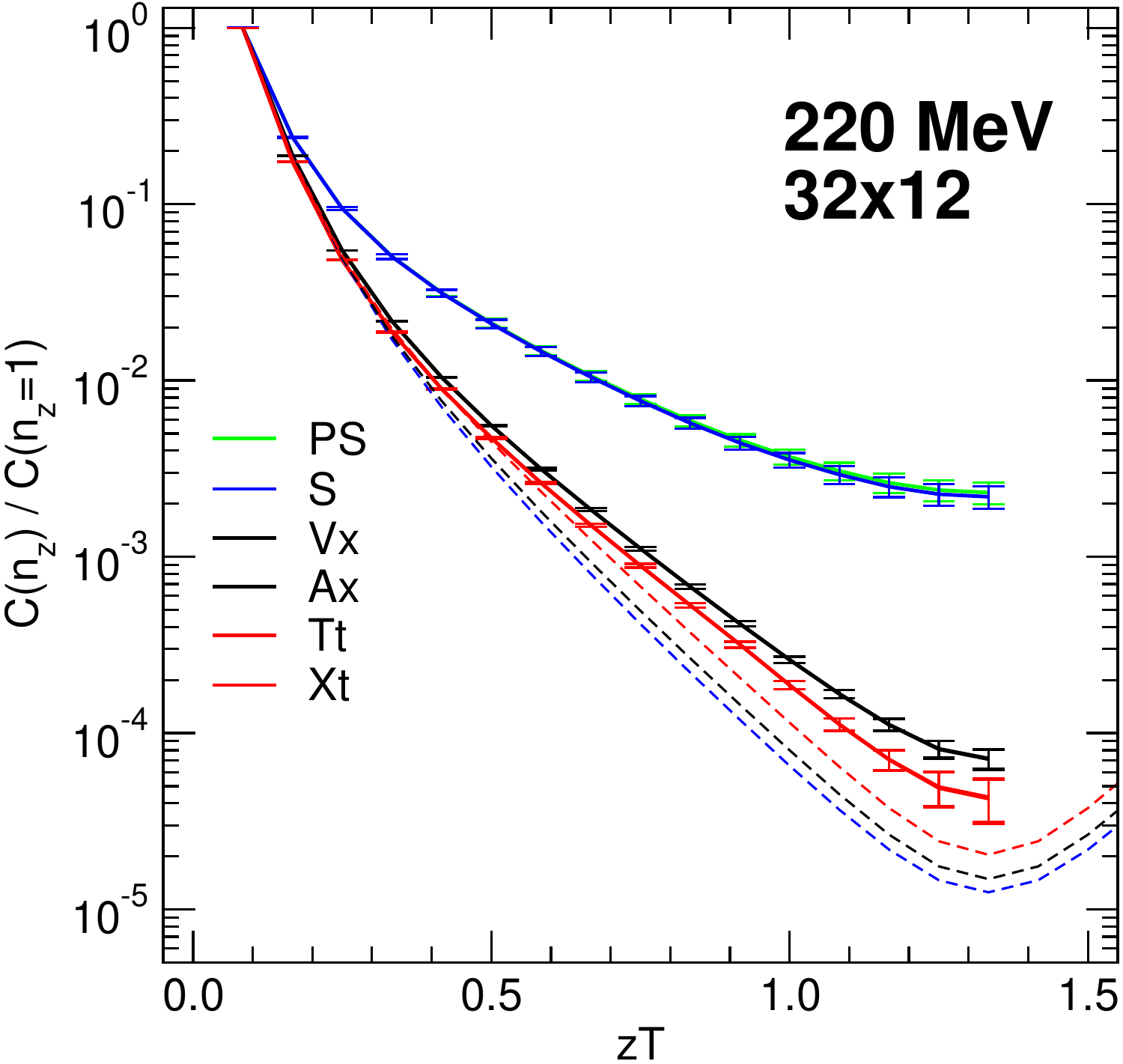}
\includegraphics[width=0.32\textwidth]{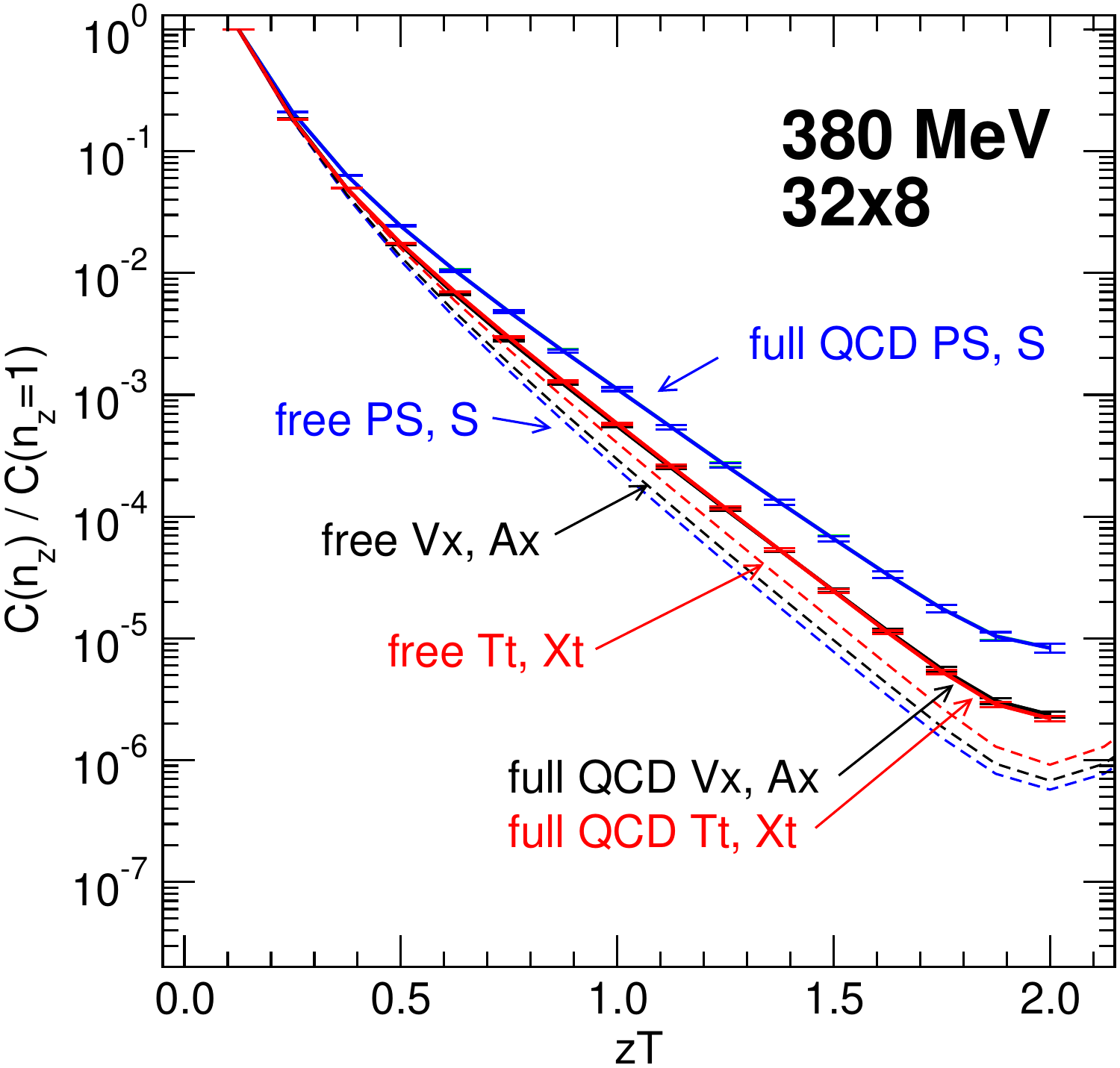}
\includegraphics[width=0.32\textwidth]{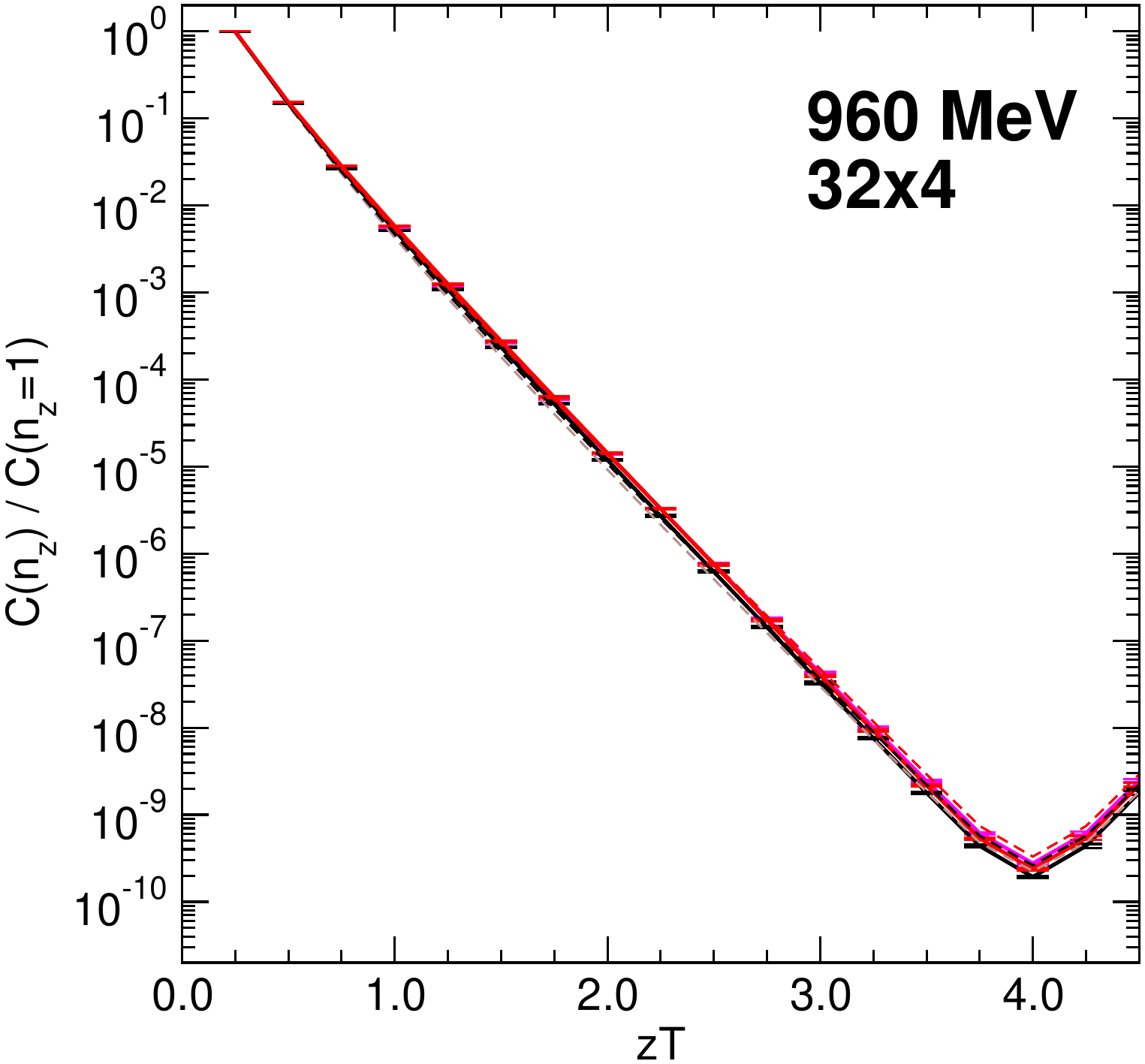}
\caption[]{Spatial correlation functions of the $E_1,E_2$ multiplets of the $SU(4)$ chiral spin symmetry, as a function of temperature
and in comparison to free quark correlators (dashed lines) \cite{Rohrhofer:2019qwq}.}
\label{fig:chi_spin}
\end{figure}

The realisation of these symmetries has recently been tested with spatial \cite{Rohrhofer:2019qwq} and 
temporal \cite{Rohrhofer:2019qal} correlation functions. \fig\ref{fig:chi_spin} shows some examples of spatial correlators,
\beq
C_\Gamma(n_z)=\sum_{n_x,n_y,n_\tau} \langle O_\Gamma(n_x,n_y,n_z,n_\tau)O_\Gamma(\mathbf{0},0)\rangle\;,
\eeq
with quantum numbers specified by $\Gamma$,
evaluated on JLQCD configurations with $N_f=2$ domain wall fermions 
with physical light quark masses from $N_\tau=4,6,8,12$ lattices.            
At $T=220$ MeV, i.e.~above the chiral crossover, a near-degeneracy pattern of different quantum number channels is observed, 
which is consistent with the multiplets of the $SU(4)$-symmetry worked out in \cite{Rohrhofer:2019qwq}.  
As the temperature is increased to $380$ MeV, these multiplets move closer to each other, while in both cases they
differ clearly from the leading-order perturbative pattern expected for free quarks. Finally, as the temperature approaches $\sim 1$~GeV,
the different multiplets fall on top of each other and approach those of free quark correlators, signalling restoration of the full chiral 
symmetry.  

This has implications for the active degrees of freedom in the temperature range just above the chiral crossover.
The nearly intact multiplet structure implies suppression of colour-magnetic interactions, and the authors interpret the
active degrees of freedom as chiral quarks bound to colour singlets by colour-electric strings. For this reason they term this regime,
which has more resemblance to a hadron gas than to a plasma, a ``stringy fluid''.

This finding based on an emergent chiral symmetry is in fact consistent with 
earlier studies distinguishing between colour-electric and -magnetic 
degrees of freedom by the discrete transformation of Euclidan time reversal (or $CT$ in Minkowski time) \cite{Arnold:1995bh}. 
The lowest screening masses 
in the framework of dimensionally reduced QCD, valid for temperatures above the crossover, and thus the dominant degrees of freedom,
correspond to colour-electric operators, with
colour magnetic ones contributing to excited states only \cite{Hart:2000ha}. 
This is contrary to the perturbative ordering of electric and magnetic scales,
$\sim gT, g^2T$, respectively, which is only realised at much higher temperatures. Finally, the picture of a hadron-like stringy fluid
is also consistent with sequential melting scneario of heavy quarkonia at $T>T_{pc}$.

\subsection{The crossover at small baryon densities}

\begin{figure}[t]
\centering
\vspace*{-0.5cm}
\includegraphics[width=0.68\textwidth]{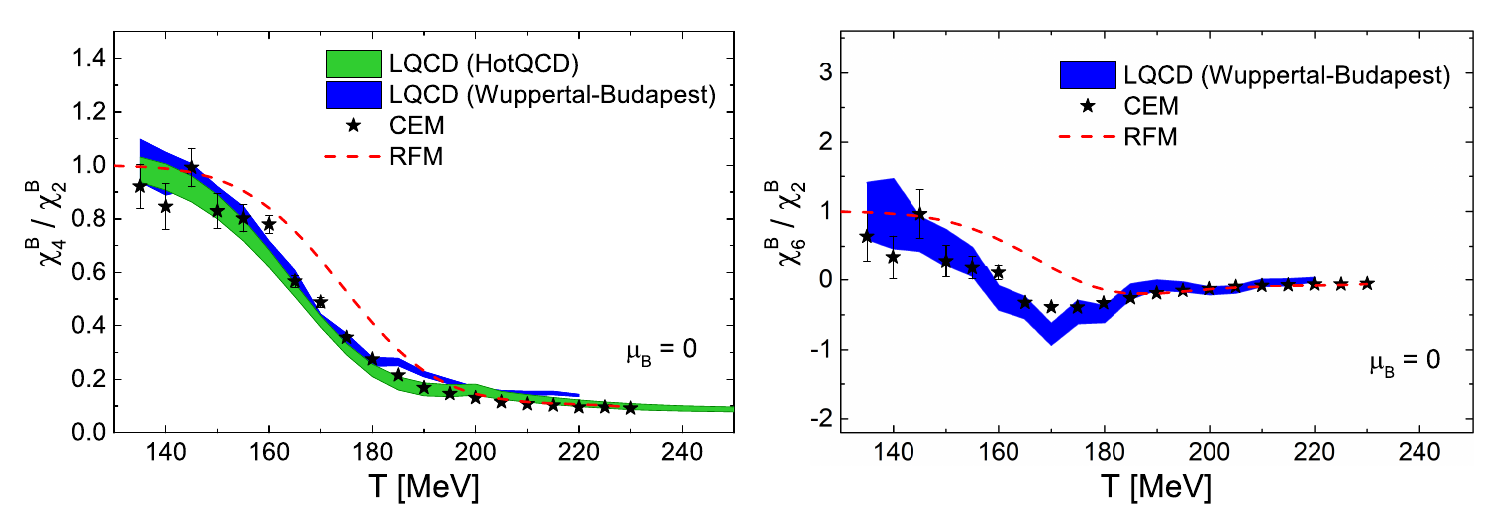}
\includegraphics[width=0.31\textwidth]{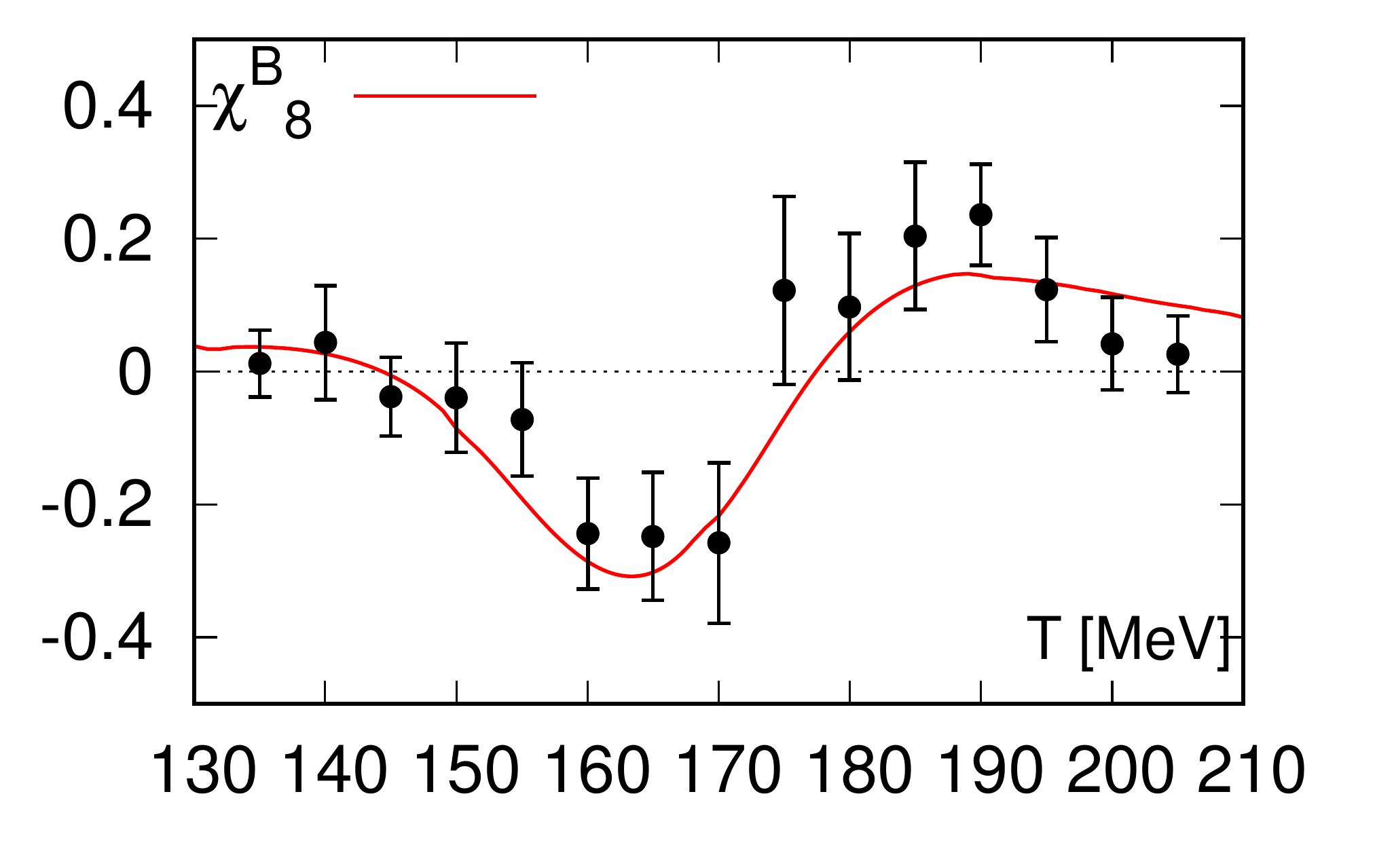}
\caption[]{Baryon number fluctuations $\chi^B_4,\chi^B_6$ from the lattice in comparison with CEM, RFM 
models \cite{Vovchenko:2019vsm} (left, middle), and $\chi^B_8$
fitted with a polynomial model without criticality \cite{Fodor:2019ogq} (right).}
\label{fig:cemtest}
\end{figure}

There are three methods that have been used so far to extract information about
the phase structure at the physical point for small baryon density. All of them introduce some approximation which can be controlled 
as long as $\mu/T\lsim 1$: i) reweighting \cite{Fodor:2001au}, ii) Taylor expansion in $\mu/T$ \cite{Allton:2002zi}
and iii) anlaytic continuation from imaginary chemical potential \cite{deForcrand:2002hgr,DElia:2002tig}. 
When the QCD pressure is expressed as a series in baryon chemical potential,                
\beq
\frac{p(T,\mu_B)}{T^4}=\frac{p(T,0)}{T^4}+\sum_{n=1}^\infty \frac{1}{2n!} \chi^B_{2n}(T) \left(\frac{\mu_B}{T}\right)^{2n}\;,\quad 
 \chi^B_{2n}(T)=\frac{\partial^{2n} (\frac{p}{T^4})}{\partial (\frac{\mu_B}{T})^{2n}}\Big|_{\mu_B=0}\;,
 \label{eq:press}
\eeq
the Taylor coefficients are the baryon number fluctuations evaluated at zero density, which can also be computed by fitting
to untruncated results at imaginary $\mu_B$, thus permitting full control of the systematics between ii) and iii).
They are presently known up to $2n=8$ on $N_\tau=16$ lattices, \fig\ref{fig:cemtest}, and in principle also observable experimentally.
For a review of the equation of state relating to heavy ion phenomenology, see \cite{Ratti:2019tvj}. Note also, that this regime
appears now accessible by complex Langevin simulations without series expansion, 
albeit not yet for physical quark masses \cite{Sexty:2019vqx}.
From the susceptibility of an appropriately normalised chiral condensate follows the pseudo-critical temperature, similarly as a power series.
The latest continuum extrapolated results are
\beq
 \frac{T_{pc}(\mu_B)}{T_{pc}(0)}=1+\kappa_2\left(\frac{\mu_B}{T}\right)^2+\ldots,\quad
\kappa_2= \left\{
 \begin{array}{llc}
0.0135(20) & \mbox{imag.}\, \mu, \mbox{stout-sm. stag.}& \cite{Bonati:2018wdn} \\
0.0145(25) & \mbox{Taylor, stout-sm. stag.} &  \cite{Bonati:2018wdn,Bonati:2018nut} \\
0.016(5) & \mbox{Taylor, HISQ} & \cite{Bazavov:2018mes}\\
 \end{array}
 \right.
 \eeq
with $T_{pc}(0)=156.5(1.5)$ MeV \cite{Bazavov:2018mes}, and the sub-leading term insignificant to current accuracy.

\subsection{The radius of convergence}

If a function with a given domain of analyticity in its complex argument is expanded in a power series, 
the radius of convergence specifies the distance between the expansion point and the nearest singularity. 
This implies that the location $(T_c,\mu_B^c)$ of a non-analytic QCD phase transition constitutes an upper bound on the 
radius of convergence of the pressure series (\ref{eq:press}). It has been argued in the literature that 
this can be turned around in order to search for a critical point: if a finite
radius of convergence can be extracted from the pressure series for real parameter values, it should signal a phase transition.
The standard estimator used in the literature is the ratio test of consecutive coefficients, whose extrapolation to infinite order
yields the radius of convergence,
\beq
r=\lim_{n\rightarrow \infty} r_{2n}\;,\quad r_{2n}=\left|\frac{2n(2n-1)\chi^B_{2n}}{\chi^B_{2n+2}}\right|\;.
\eeq
In practice, only the first few coefficients are available. Nevertheless, several constraints on a critical endpoint have been based on $r_{2n}$ 
and published, a recent compilation can be found in \cite{Bazavov:2017dus}. 

However, it has recently become clear in model studies that the ratio
estimator is inappropriate for the case at hand.
As an example, consider the fugacity expansion of baryon number density. At imaginary chemical potential, this 
is just a Fourier series whose coefficients can be computed on the lattice without sign problem,  
\beq
\frac{n_B}{T^3}|_{\mu_B=i\theta_B T=}=i\sum_k b_k(T) \sin(k\mu_B/T)\;,\quad
b_k(T)=\frac{2}{\pi}\int_0^\pi d\theta_B\; {\rm Im}(\frac{n_B(T,i\theta_BT)}{T^3}\sin(k\theta_B)\;.
\eeq  
In \cite{Vovchenko:2017gkg} a cluster expansion model (CEM) was proposed, which takes the first two coefficients as input
from a lattice calculation \cite{Vovchenko:2017xad}, and expresses all higher coefficients in terms of these,
\beq
b_k(T)=\alpha_k^{SB}\frac{[b_2(T)]^{k-2}}{[b_1(T)]^{k-1}}\;,\quad k=3,4,\ldots
\eeq
The $\alpha_k^{SB}$ are $T$-independent and fixed to reproduce the Stefan-Boltzmann limit.
Hence, this includes two-body interactions only, corresponding to NLO in a virial expansion, which should be valid at sufficiently
high temperatures and low densities. Of course, modelling higher coefficients in terms of the lower ones is not unique, an 
alternative is provided by the rational function model (RMF) \cite{Almasi:2019bvl}.
The model now predicts the coefficients $b_{k\geq3}$ or any of the baryon number susceptibilities,
a closed expression to all orders as a polylogarithm is also available \cite{Vovchenko:2019vsm}. 
All existing lattice data are reproduced surprisingly accurately, as the examples in \fig\ref{fig:cemtest} show.
\begin{figure}[t]
\centering
\vspace*{-0.5cm}
\includegraphics[width=0.47\textwidth]{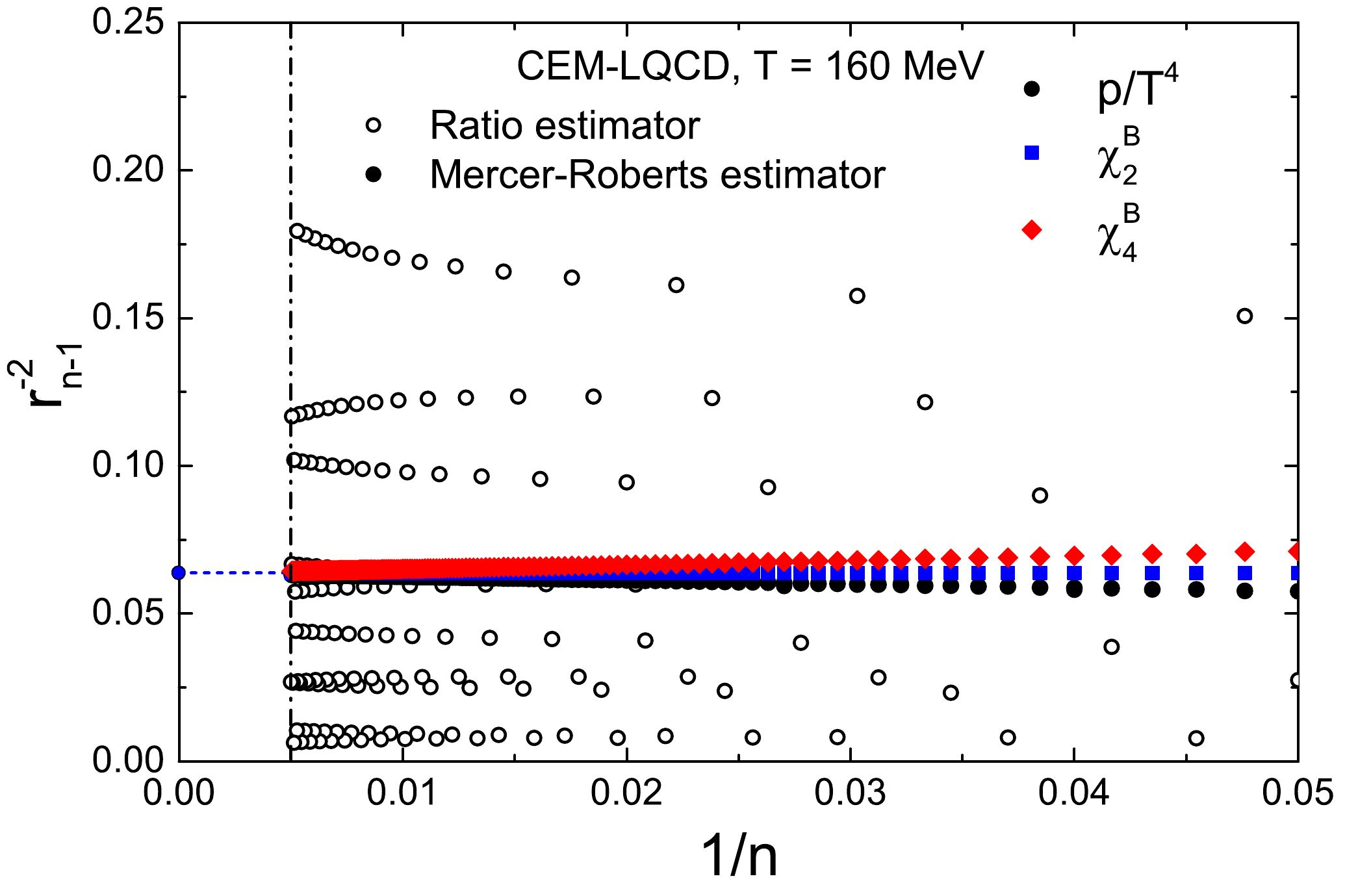}\hspace*{0.5cm}
\includegraphics[width=0.5\textwidth]{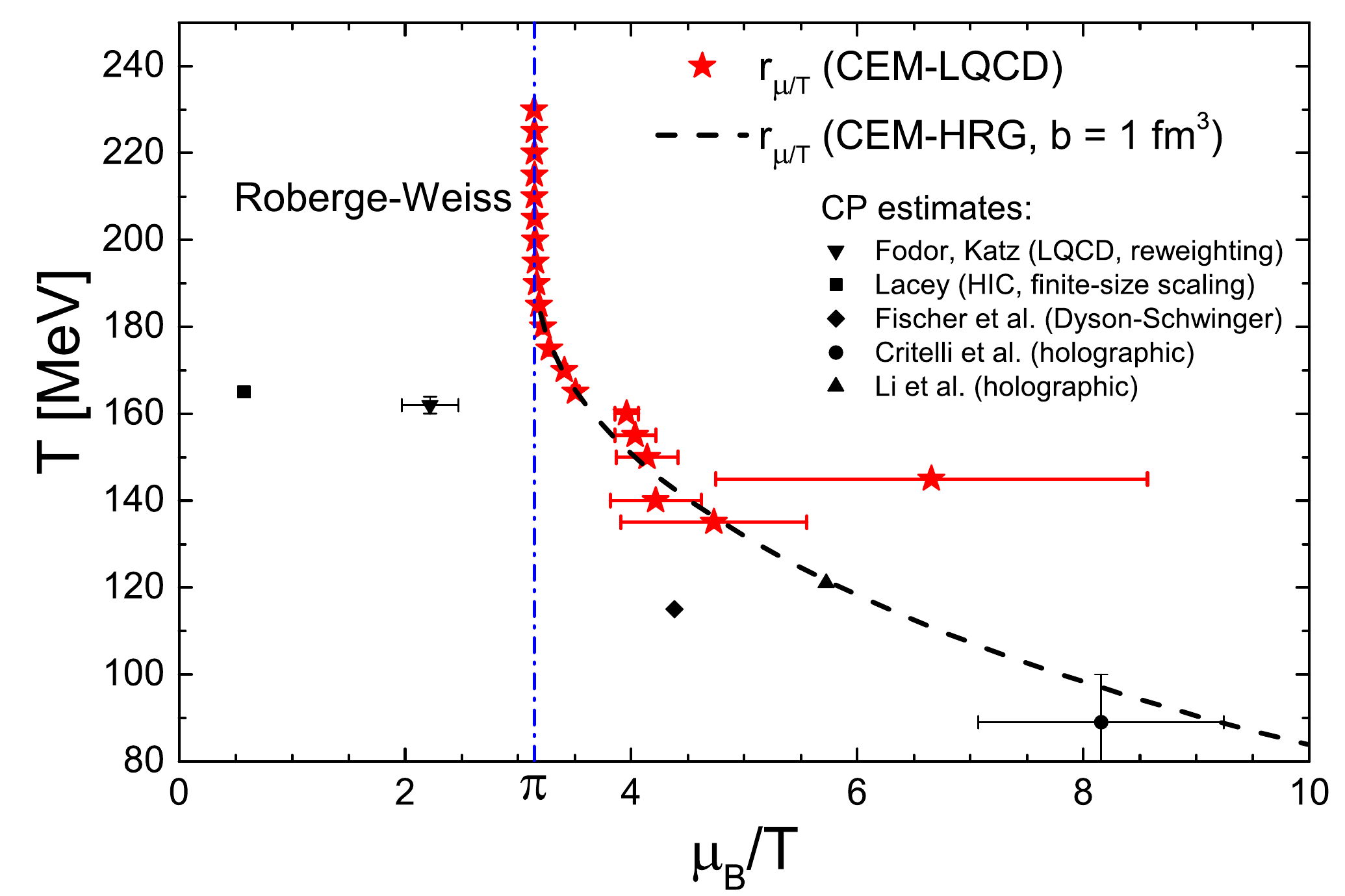}
\caption[]{Left: Comparison of estimators for the radius of convergence in a Cluster Expansion Model. Right:
The CEM radius of convergence as a function of $T$ predicts the RW-transition at imaginary $\mu_B$ \cite{Vovchenko:2017gkg}.}
\label{fig:cem_rad}
\end{figure}

With all coefficients of the fugacity expansion available, one can study the radius of convergence of CEM, \fig\ref{fig:cem_rad}. 
The ratio estimator fails to converge, because of the irregular signs of higher order coefficients
(it works for equal or alternating signs). On the other hand, the Mercer-Roberts
estimator, which uses three consecutive coefficients of a series \cite{MercerRoberts},
\beq
r_n=\left|\frac{c_{n+1}c_{n-1}-c_n^2}{c_{n+2}c_n-c_{n+1}^2}\right|^{1/4}\;,
\eeq
converges and extrapolates to a unique radius of convergence, independent of the observable used. 
 \fig\ref{fig:cem_rad} (right) shows the result for various temperatures, which intriguingly predicts the Roberge-Weiss transition in the
direction of imaginary chemical potential. Conversely, this implies that the CEM  has no
phase transition for real $\mu_B\leq \pi T$. Of course, this is a model and does {\it not} exclude a QCD critical point in this range. (For 
an application of the RFM to a chiral model which has a phase transition, see \cite{Almasi:2019bvl}.) However, the analysis {\it does} 
imply that there is no sign of criticality in the presently available lattice data at zero or imaginary chemical potential (see 
also \fig\ref{fig:cemtest} (right)).

A more general study investigates how the singularity structure of QCD is reflected in Lee-Yang zeroes and
the radius of convergence \cite{Giordano:2019slo}. 
It concludes that the ratio test fails in any finite volume and also advocates the Mercer-Roberts 
as well as the Cauchy-Hadamard estimators. Improved versions are constructed for these estimators 
with enlarged sensitivity to the closest singularity, which are able to pick up the known phase transition of $N_f=4$ staggered 
QCD on a $N_\tau=4$ lattice with less than ten coefficients. 

Having to go term by term in an expansion can be avoided, if the Lee-Yang zero closest to the 
origin can be determined directly. Using reweighting, this was the strategy employed in the first prediction of a 
critical point on $N_\tau=4$ lattices using unimproved rooted staggered fermions \cite{Fodor:2001pe}. 
However, a new investigation points out that, 
for this discretisation, the closest Lee-Yang zero is caused by a spectral gap between the unrooted tastes 
rather than by a phase transition \cite{Giordano:2019gev}. A new definition of the rooted staggered determinant 
at finite $\mu$ is suggested, which avoids these artificial non-analyticies. Application to  the stout-smeared action on $N_\tau=4$, again using
reweighting, shows the closest singularity to be off the real axis, pushing a possible phase transition beyond $\mu_B\gsim 2T$. 

\subsection{Finite isospin density}

\begin{figure}[t]
\centering
\vspace*{-0.5cm}
\includegraphics[width=0.33\textwidth]{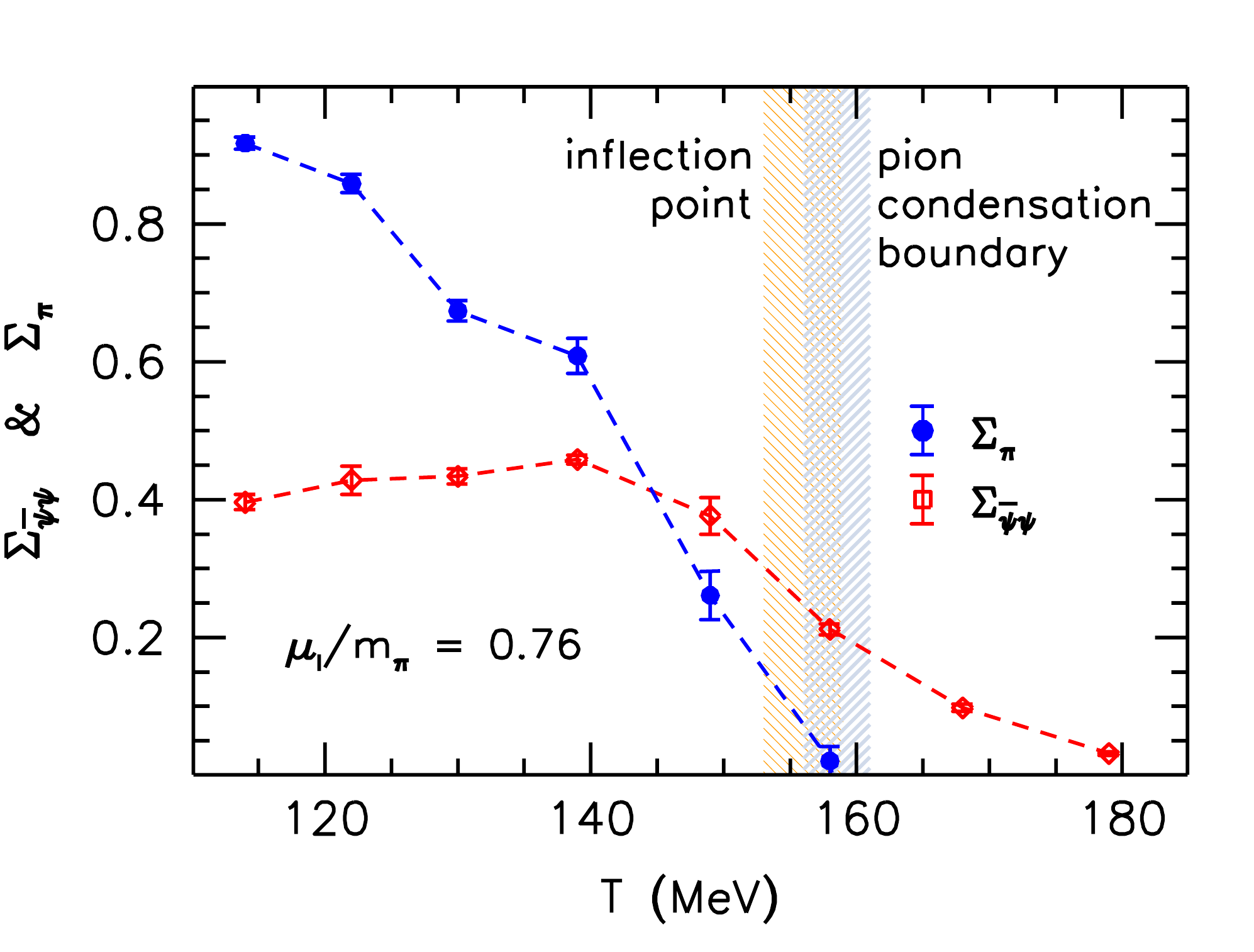}
\includegraphics[width=0.32\textwidth]{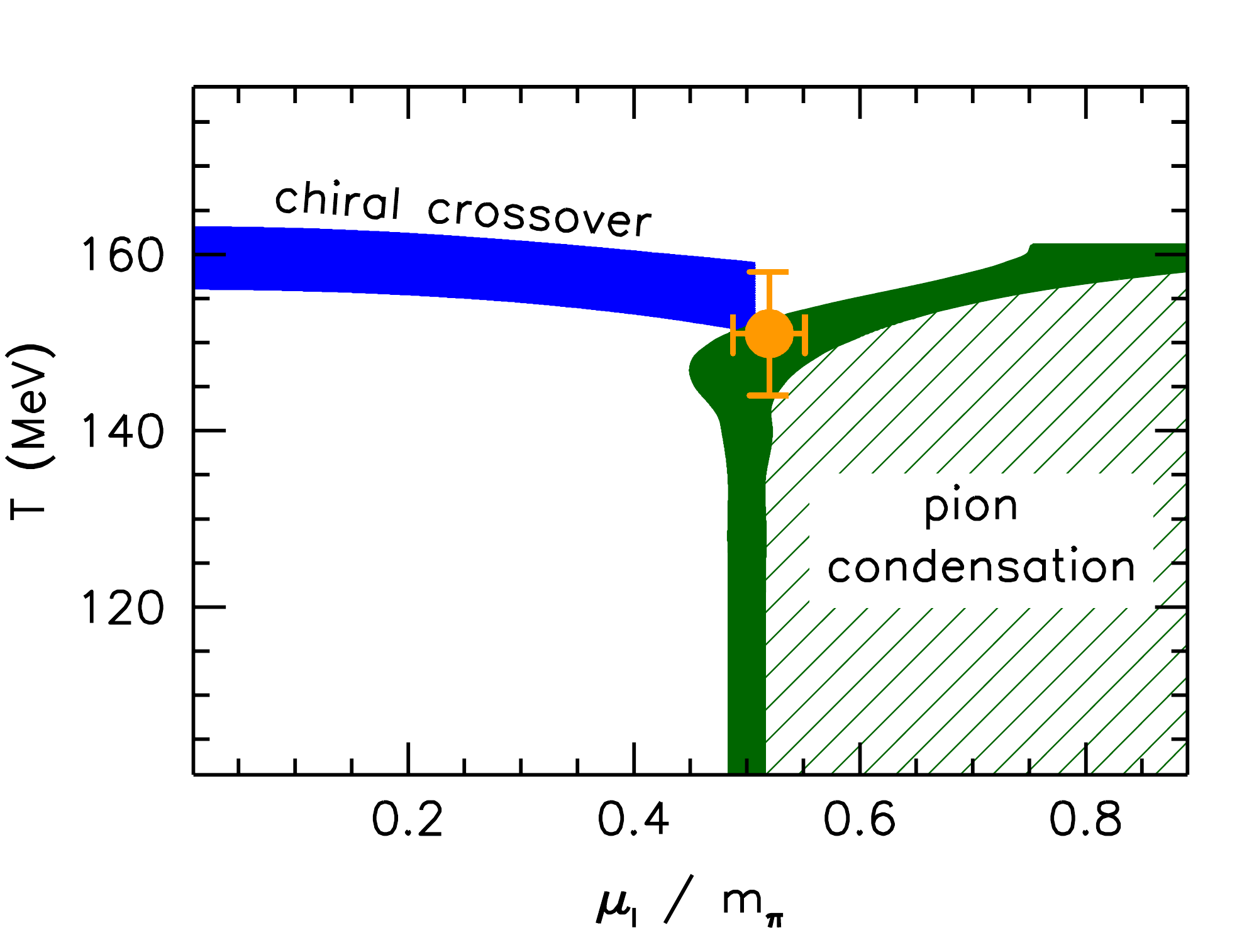}
\includegraphics[width=0.33\textwidth]{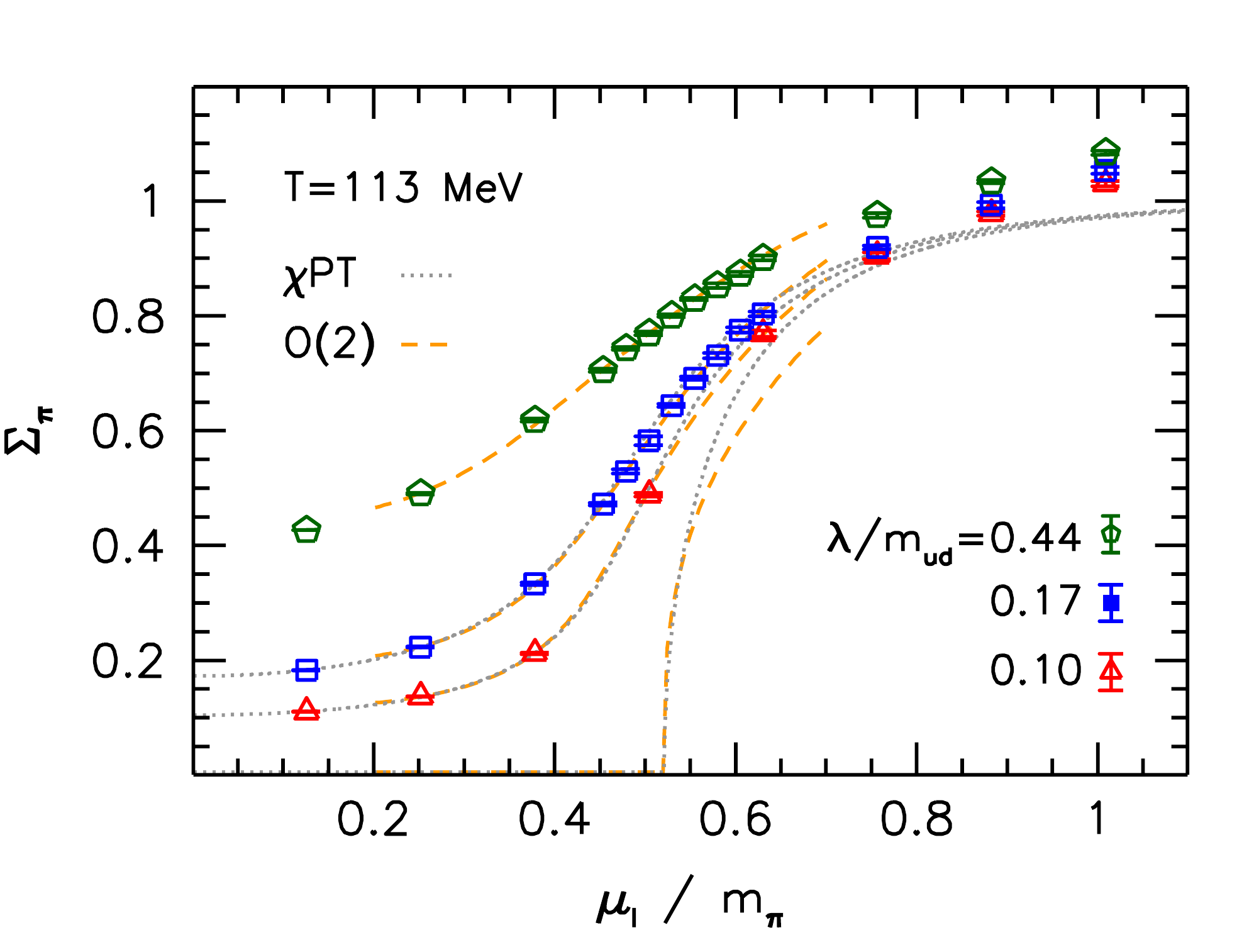}
\caption[]{Left: Chiral and isospin condensated as a function of $T$. Middle: Phase diagram at finite
isospin chemical potential. Right: Scaling of the pion condensate
as a signal for a second-order transition \cite{Brandt:2017oyy}.}
\label{fig:iso}
\end{figure}
It was noted long ago that QCD at finite isospin density does not have a sign problem and can be simulated 
directly \cite{Alford:1998sd}. Finite isospin is also physically relevant for neutron stars or 
the early universe with a lepton asymmetry. Consider the (degenerate) light quark action in the form
\beq
S=\bar{\psi}\left(\gamma_\mu(\partial_\mu+iA_\mu)+ m_{ud}+\mu_I\gamma_4\tau_3+i\lambda\gamma_5\tau_2\right)\psi\;,\quad
\mu_I=(\mu_u-\mu_d)/2 \;.
\eeq
A chemical potential for isospsin is realised by having chemical potentials of opposite sign for the $u$- and $d$-quarks.
For $\mu_I=0$, the action is invariant under a $SU(2)_V\times U(1)_V$ associated with isospin and baryon number, which gets
broken to $U(1)_{\tau_3}\times U(1)_V$ by $\mu_I\neq 0$. In this case there is a further spontaneous symmety breaking leaving
only the baryon $U(1)_V$, signalled by a non-vanishing expectation value for the charged pions, 
\beq
\langle \pi^\pm\rangle=\langle\bar{\psi}\gamma_5\tau_{1,2}\psi\rangle \;,
\eeq
which then correspond to Goldstone modes. The $\lambda$-term in the action introduces an explicit breaking of the remaining 
symmetry and is necessary for simulation purposes only, in order to pick one of the degenerate vacua. 
For physical results, simulations have to be 
extrapolated to $\lambda\rightarrow 0$, a task quite similar to approaching the chiral limit at $\mu=0$, and hence
difficult. 

After several exploratory studies (e.g.~\cite{Kogut:2002zg,deForcrand:2007uz,Cea:2012vi}) new calculations 
are
for $N_f=2+1$ quarks with physical masses using a stout-smeared staggered action, on lattices with $N_\tau=6,8,10,12$, 
followed by a continuum extrapolation. The $\lambda\rightarrow 0$ extrapolation is done with a similar reweighting technique applied to 
a singular value representation of the pion condensate, as for the chiral condensate mentioned in 
section \ref{sec:scale} \cite{Endrodi:2018xto}.
The order parameters used to characterise the phase structure are the chiral and pion condensates,
\beq
\Sigma_\pi=\frac{m_{ud}}{m_\pi^2f_\pi^2}\langle \pi^\pm\rangle\;, \quad
\Sigma_{\bar{\psi}\psi}=\frac{m_{ud}}{m_\pi^2f_\pi^2}(\langle \bar{\psi}\psi\rangle_{T,\mu_I}-\langle \bar{\psi}\psi\rangle_{0,0})+1\;,
\eeq
which are shown as a function of $T$ for a specific choice of $\mu_I$ in \fig\ref{fig:iso} (left). Varying both thermodynamic parameters,
the phase boundaries have been mapped out leading to the phase diagram in   \fig\ref{fig:iso} (middle). 
For the transition in the $\mu_I$-direction, the extrapolation
$\lambda\rightarrow 0$ is consistent with $O(2)$ scaling of the pion condensate, identifying the transition to be second-order.
Prospects to reweight in $\mu_B$ using finite isospin as a point of departure are discussed in \cite{Schmalz}.

\section{Larger baryon density via effective lattice theories}

For the cold and dense regime, $\mu_B/T\gg 1$, where the sign problem is strongest, no genuine methods are available.  
Nevertheless, some progress towards at least qualitative physics has been made over
the last few years by means of effective theories. The general idea is to split the problem in two parts: first
derive an effective theory by expansion methods in some small parameter. By this step some degrees of freedom 
have already been integrated over, such that the sign problem for the resulting effective theory is milder than the original one.
In a second step the effective theory is solved by flux representations
simulated by a worm algorithm, complex Langevin simulations or analytic series expansion methods. 
\begin{figure}[t]
\vspace*{-0.5cm}
\centering
\includegraphics[width=0.45\textwidth]{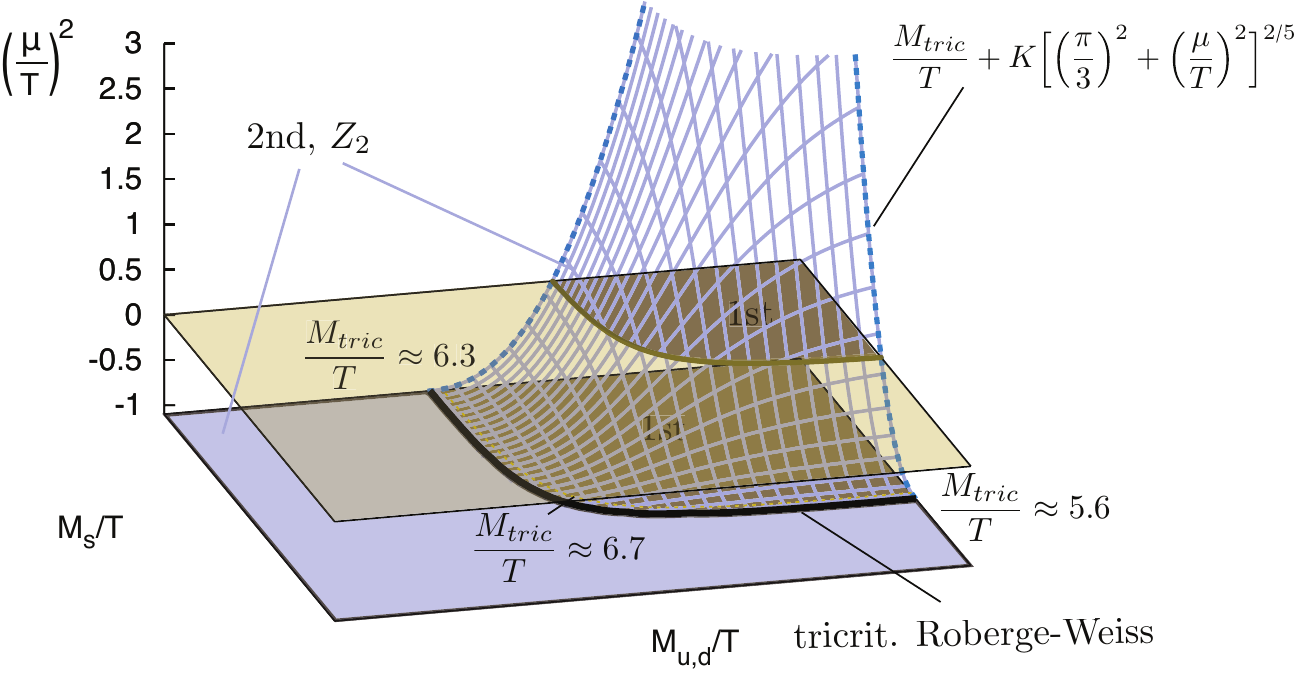}\hspace*{1cm}
\includegraphics[width=0.4\textwidth]{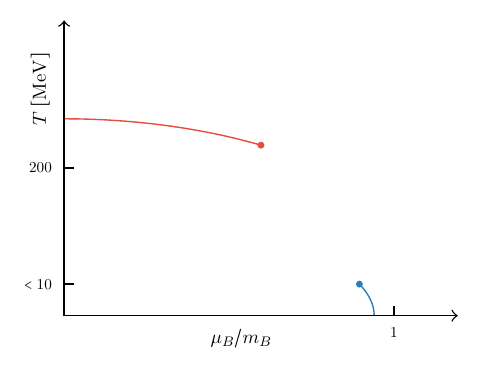}
\caption[]{ Left: 
Heavy mass corner of the Columbia plot, computed in an effective theory \cite{Fromm:2011qi}.
Right: Schematic phase diagram for QCD with heavy quarks.}
\label{fig:efft_pd}
\end{figure}

Two types of effective degrees of freedom arise naturally, depending on the integration order,
\beq
Z=\int DUD\bar{\psi}D\psi\;e^{-S_{QCD}[U,\bar{\psi},\psi]}=\int DU_0\;e^{-S_{eff}[U_0]} 
=\int D\bar{\psi}D\psi\;\;e^{-S_{eff}[\bar{\psi},\psi]}\;.
\eeq
In the first case, fermions are integrated over as well as all spatial link variables, leaving a theory of temporal links only, which
on a periodic lattice can be expressed by Polyakov loops. In the second case, all gauge links are integrated, leaving a fermionic
effective theory in terms of mesons and baryons, because of gauge invariance. Note that both representations are
perfectly equivalent to QCD. Because of the truncations involved in doing the integrations analytically, this equivalence is reduced
to specific parameter regions, where the approximations hold.

\subsection{Effective theory for heavy quarks}

The representation of QCD in terms of Polyakov loops was developed to characterise the thermal transition in pure gauge 
theories \cite{Polonyi:1982wz,Svetitsky:1982gs} and has been extended to include fermions. Starting point is Wilson's lattice
formulation. The effective theory is then derived by a combined expansion in the fundamental character coefficient, which is
a known function of the gauge coupling and always smaller than one for finite $\beta$, and the hopping parameter,
\beq
u(\beta)=\frac{\beta}{18}+\frac{\beta^2}{216}+\ldots < 1\;,\quad \frac{1}{8}<\left(\kappa=\frac{1}{2am+8}\right)<\frac{1}{4}\;.
\eeq
With the euclidean time extent now absorbed in the remaining temporal Wilson line variables $W(\bx)$, 
the result is an effectively 3d theory resembling
a continuous spin model \cite{Langelage:2010yr,Fromm:2011qi},
\bea
\label{zpt}
Z&=&\int DW\prod_{<\bx, \by>}\left[1+\lambda(L_{\bx}L_{\by}^*+L_{\bx}^*L_{\by})\right]\\
&&\times\prod_{\bx}[1+h_1L_{\bx}+h_1^2L_{\bx}^*+h_1^3]^{2N_f}[1+\bar{h}_1L^*_{\bx}+\bar{h}_1^2L_{\bx}+\bar{h}_1^3]^{2N_f}
\nonumber\\
&&\times \prod_{<\bx, \by>}\left(1-h_{2}{\rm Tr} \frac{h_1W_{\bx}}{1+h_1W_{\bx}}{\rm Tr} \frac{h_1W_{\by}}{1+h_1W_{\by}}\right)
\left(1-h_{2}{\rm Tr} \frac{\bar{h}_1W^\dag_{\bx}}{1+\bar{h_1}W^\dag_{\bx}}{\rm Tr} \frac{\bar{h}_1W^\dag_{\by}}{1+\bar{h}_1W^\dag_{\by}}\right)
\times \ldots \;.\nn
\eea
The couplings of the effective theory, $\lambda(u,\kappa,N_\tau), h_1(u,\kappa,\mu,N_\tau), \bar{h_1}(\mu)=h_1(-\mu),h_2(u,\kappa,N_\tau)$
are (resummed) power series in the original small parameters, and can thus in turn be treated
as small expansion parameters for the effective theory.
\begin{figure}[t]
\vspace*{-0.75cm}
\centering
\includegraphics[width=0.45\textwidth]{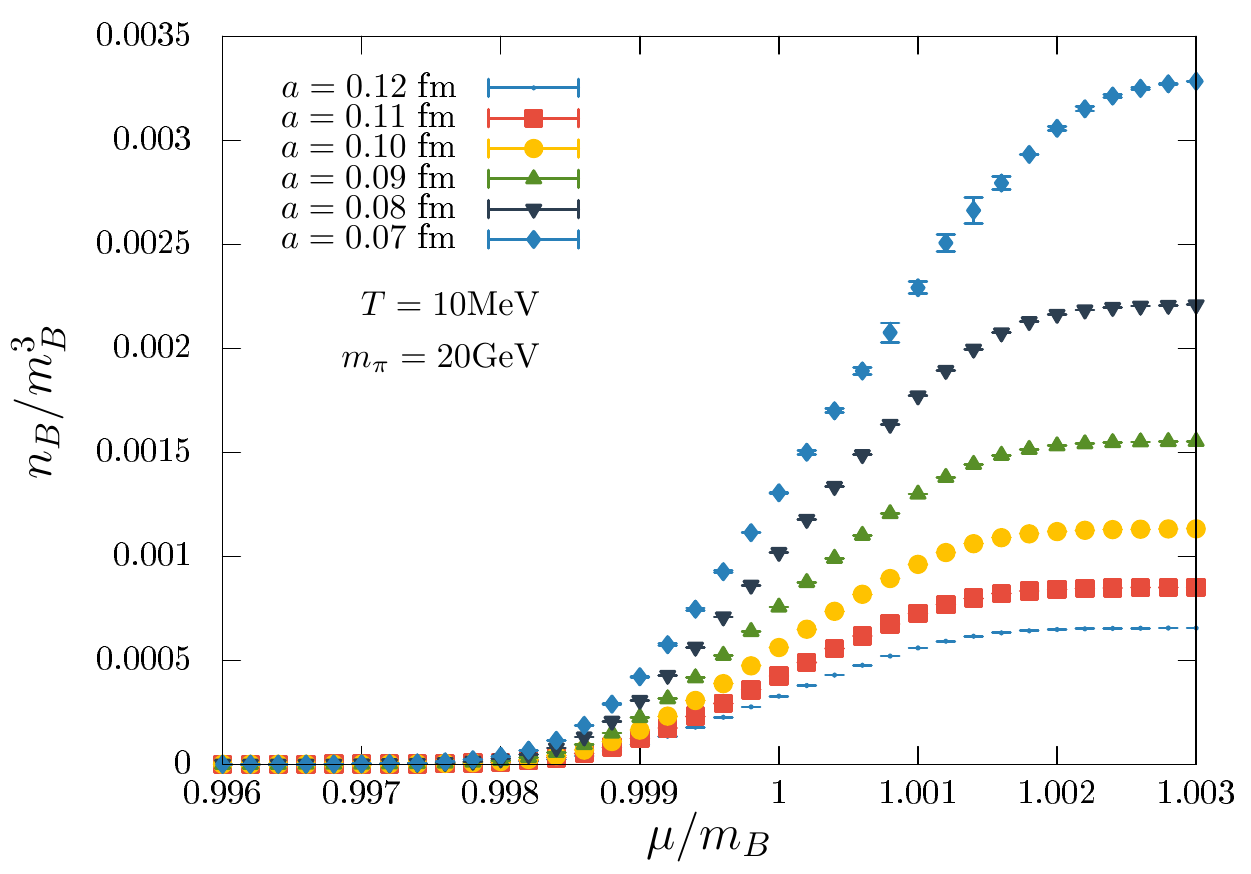}\hspace*{0.5cm}
\includegraphics[width=0.45\textwidth]{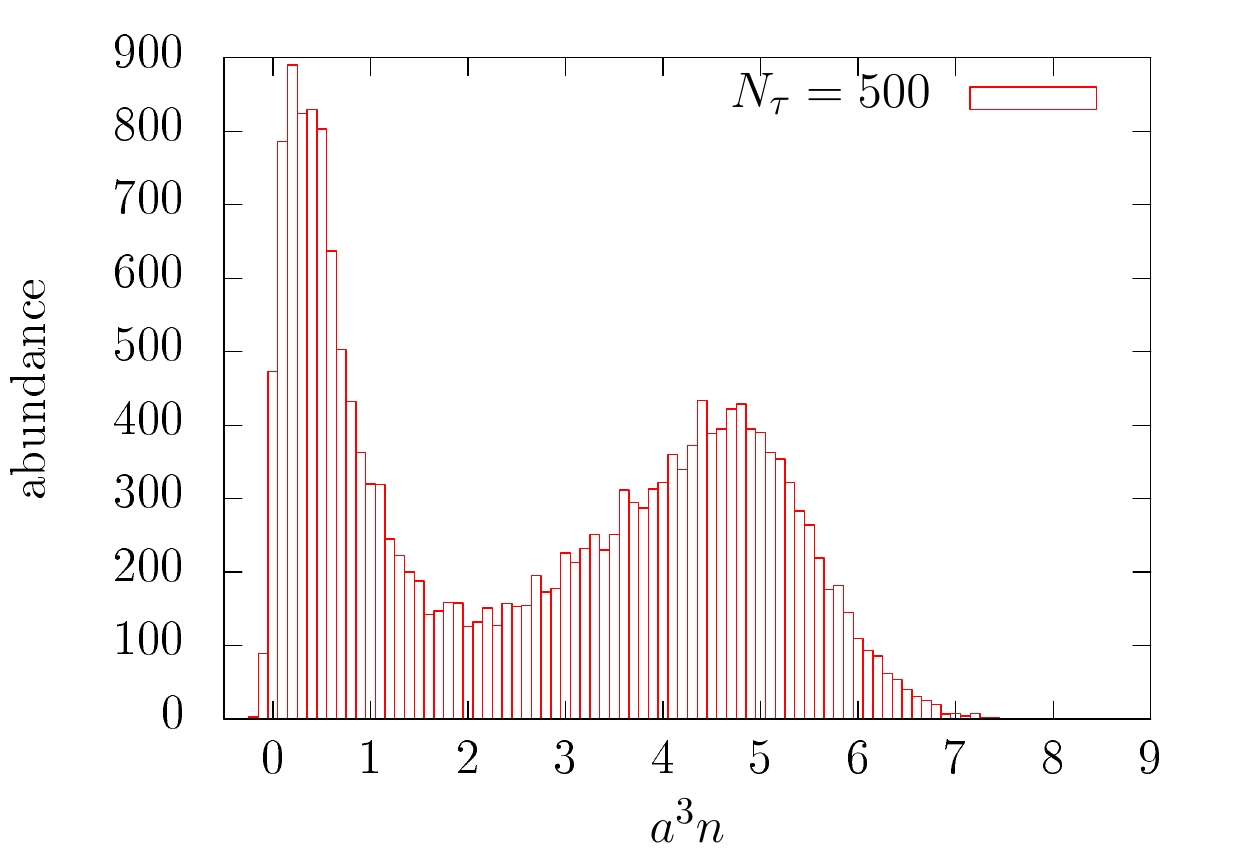}
\caption[]{Left: Onset of baryon number (crossover) for heavy quarks and different lattice spacings \cite{Glesaaen:2015vtp}.
Right: First-order onset transition for light quarks \cite{Langelage:2014vpa}.}
\label{fig:efft_dense}
\end{figure}
This is illustrated in 
\cite{quang}, where the critical coupling for the $SU(3)$ pure gauge theory  is extracted
from the effective theory by means of series expansion methods.
Comparison with the full 4d Monte Carlo result
shows agreement to better than 10\% for $N_\tau\in[2,16]$, for which continuum extrapolations are possible, thus constituting 
a completely analytic calculation of the deconfinement transition of lattice Yang-Mills theory.

When quarks are included, the heavy mass corner of the Columia plot has been simulated by means of the effective theory.
On coarse $N_\tau=4$ lattices, the critical line found in full QCD simulations at $\mu=0$ is again accurately reproduced, and the
calculation can be extended to finite $\mu$ simulating a flux representation \cite{Fromm:2011qi}. In this way the deconfinement
critical surface is known for any value of $\mu$, \fig\ref{fig:efft_pd} (left), and the phase diagram for  
heavy quarks looks like \fig\ref{fig:efft_pd} (right).

With the same methods, the cold and dense regime can also be studied and, in particular, the
onset transition to condensing baryon matter has been seen explicitly 
to various orders in the expansions \cite{Fromm:2012eb,Langelage:2014vpa,Glesaaen:2015vtp}. 
\fig\ref{fig:efft_dense} (left) shows the baryon density, featuring the ``silver blaze property'' of staying zero until $\mu_B\approx m_B$,
followed by a sudden rise to lattice saturation, which is the maximal number of quarks per lattice site allowed by the Pauli principle.
In continuum units, this lattice artefact moves to infinity by continuum extrapolation, as the figure illustrates. 
\fig\ref{fig:efft_dense} (left) shows a crossover,
whereas for light quarks simulations show a first-order transition \fig\ref{fig:efft_dense} (right) changing to crossover  
at an endpoint $T_c(m)$. Indeed,
the binding energy per baryon in the hopping expansion is found to start as $\epsilon\sim \kappa^2$ \cite{ Langelage:2014vpa},
i.e.~it decreases with growing quark mass to zero in the static limit, as one also expects from Yukawa potentials in nuclear physics.
Hence, the end point of the nuclear liquid gas transition, $T_c(m)$, decreases with mass.   

\subsection{Large $N_c$}

\begin{figure}[t]
\vspace*{-0.5cm}
\centering
\includegraphics[width=0.32\textwidth]{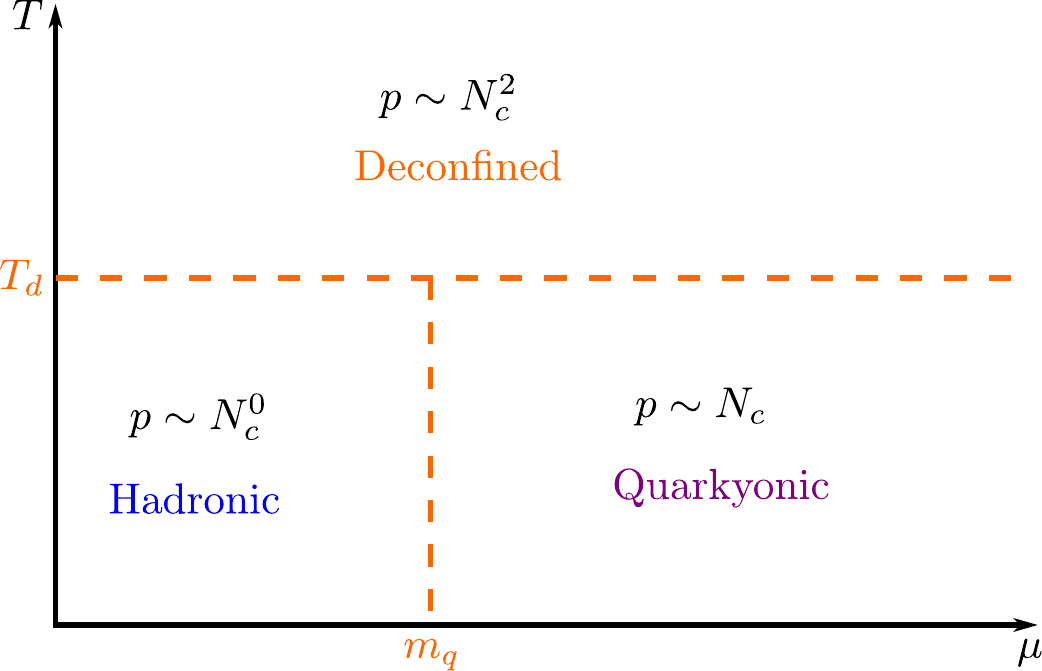}
\includegraphics[width=0.33\textwidth]{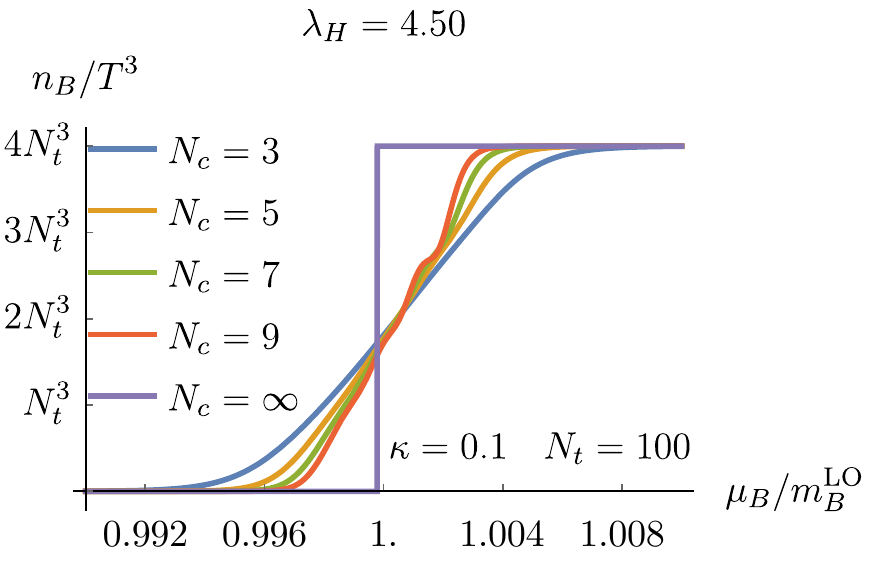}
\includegraphics[width=0.33\textwidth]{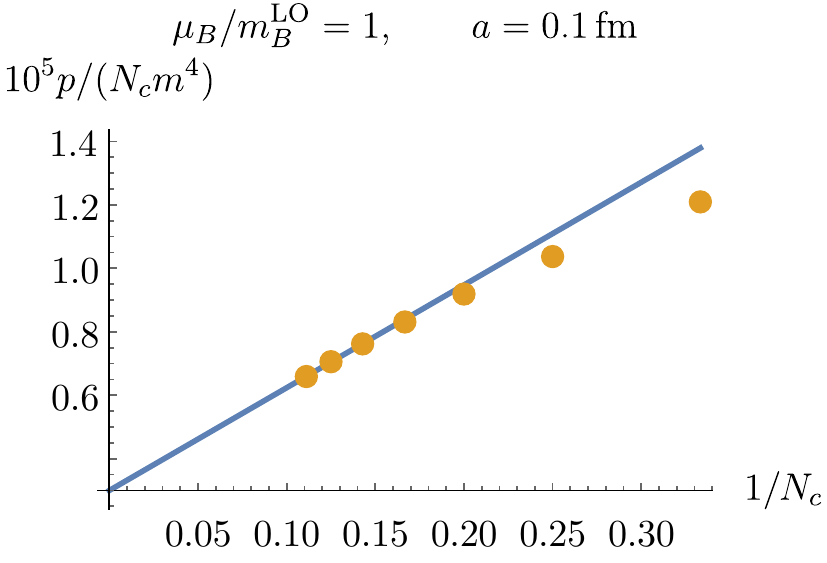}
\caption[]{Left: Phase diagram for QCD with $N_c\rightarrow \infty$ \cite{McLerran:2007qj}. Middle: The onset transition becomes first order
for large $N_c$. Right: $p\sim N_c$ for large $N_c$ in the baryon condensed region \cite{Philipsen:2019qqm}.}
\label{fig:nc}
\end{figure}

In yet another corner of QCD parameter space, interesting conjectures concerning the QCD phase structure were 
based on large $N_c$ arguments \cite{McLerran:2007qj}. In particular, the phase diagram in the large $N_c$ limit was argued to
look as in \fig\ref{fig:nc} (left). With fermion contributions suppressed, the deconfinement transition is a straight line
separating the plasma phase, where the pressure scales as $p\sim N_c^2$, from the hadron gas phase, where it scales as $p\sim N_c^0$.
In \cite{McLerran:2007qj} it is argued that at finite density there should then be a third phase with $p\sim N_c$, which was termed
quarkyonic since it shows aspects of both baryon and quark matter. In particular, the fermi sea at low temperatures is argued to be composed 
of a baryonic shell of thickness $\sim \Lambda_{QCD}$, and quark matter inside.

The effective theory of the previous section can be derived for a general number of colours~\cite{Christensen:2013xea}. 
For large $N_c$, the baryon mass is $m_B\sim N_c$, so the constituent quark mass should not matter and the cold 
and dense region for large $N_c$ is accessible to direct calculation~\cite{Philipsen:2019qqm}. 
It was found that the baryon onset transition steepens with $N_c$, to become first-order in the large $N_c$ limit, \fig\ref{fig:nc} (middle). 
Furthermore, through three orders in the hopping expansion, the pressure scales as $p\sim N_c$, suggesting this to be a property to all 
orders. This scaling is reproduced with a leading correction even for $N_c=3-9$, right after the onset transition \fig\ref{fig:nc} (right). 
The large $N_c$ phase diagram \fig\ref{fig:nc} (left) is thus continuously obtained from  \fig\ref{fig:efft_pd} (right) by increasing $N_c$.
Note also, that a lattice filling with baryon number smoothly changes from baryon matter (at the onset of condensation) 
to quark matter (at saturation) as a function of $\mu_B$, which is consistent with the picture of quarkyonic matter. 
For light quarks, there may be in addition a chiral transition.

\subsection{Effective theory for light quarks} 

\begin{figure}[t]
\centering
\vspace*{-0.3cm}
\includegraphics[width=0.5\textwidth]{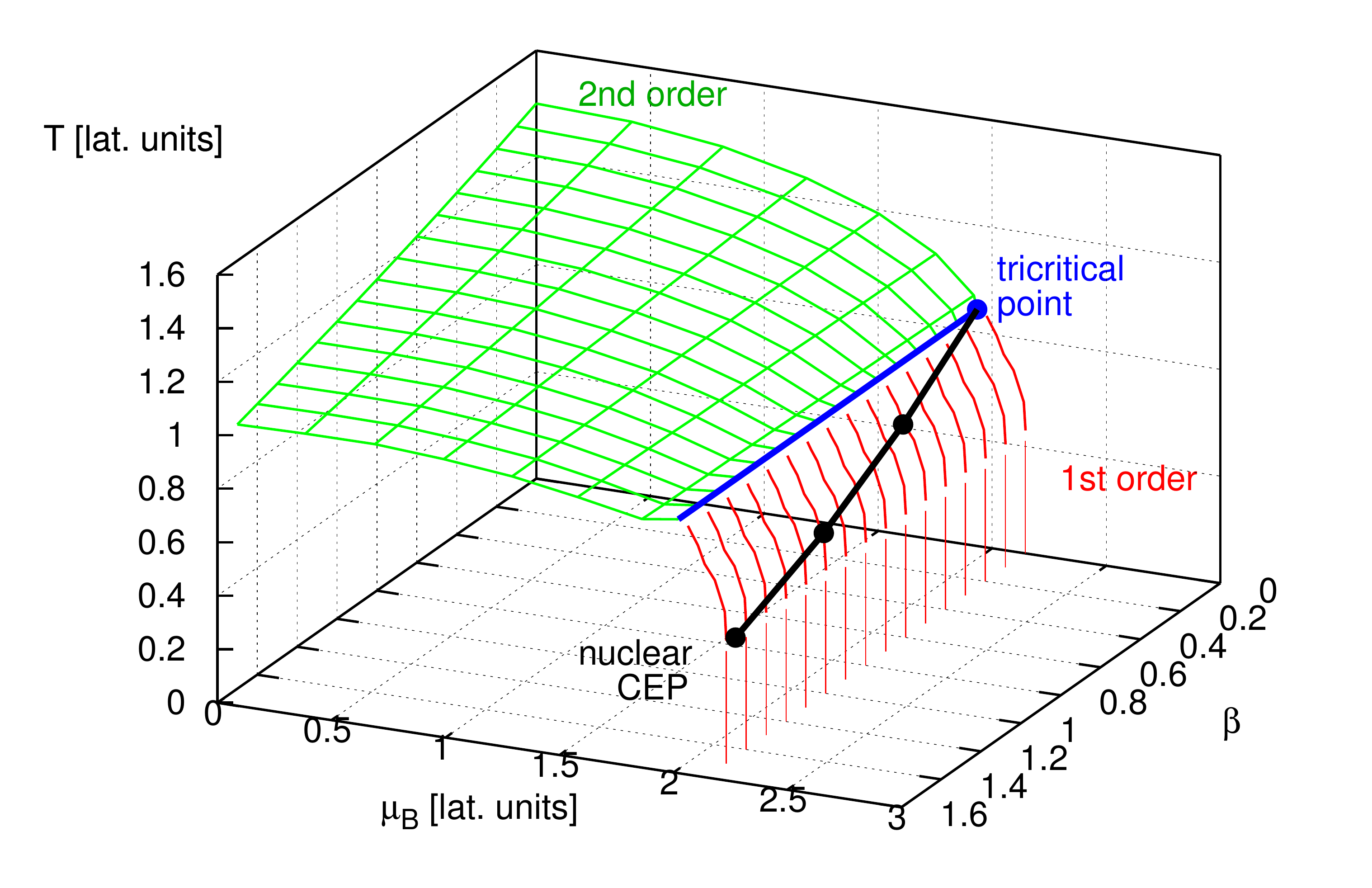}\hspace*{0.5cm}
\includegraphics[width=0.45\textwidth]{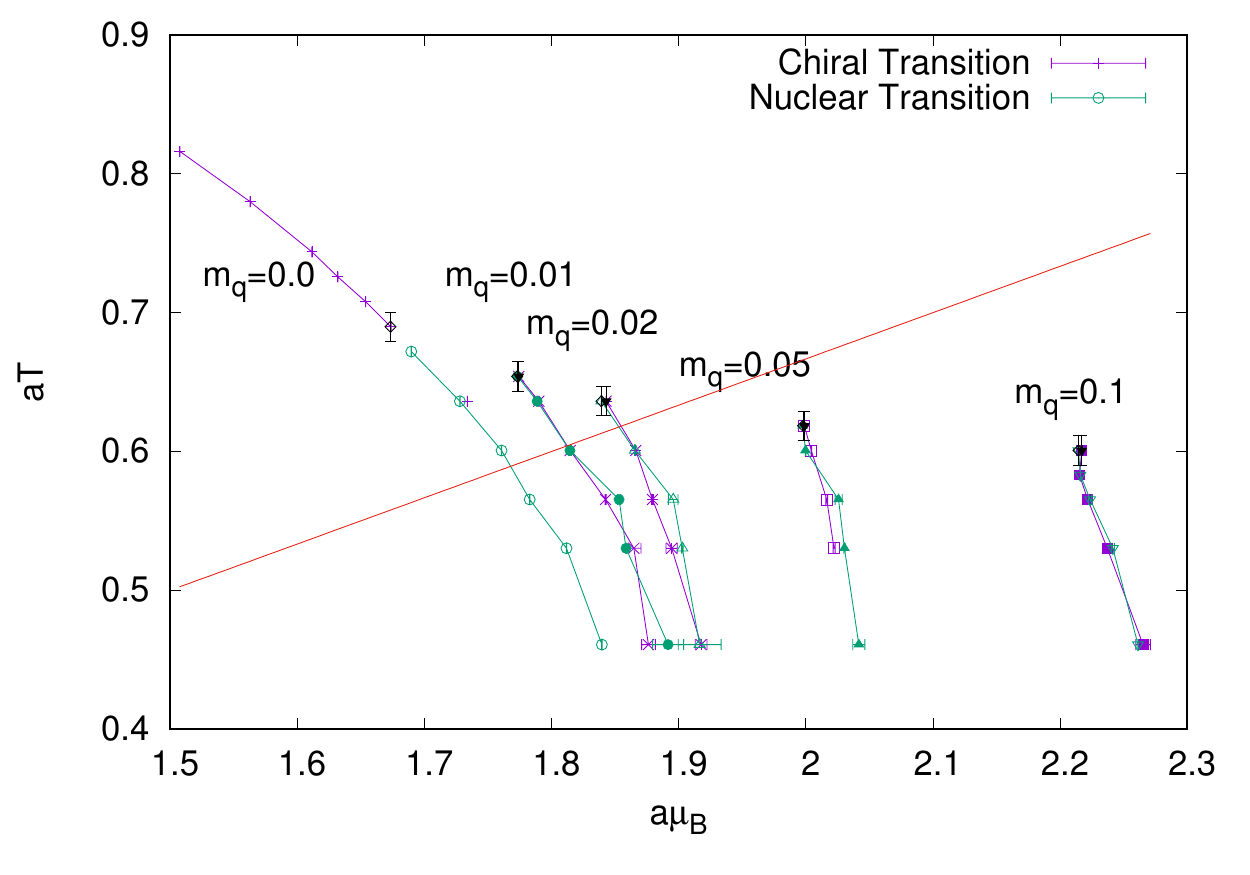}
\caption[]{Left: Chiral phase transition and onset transition to baryon condensation for $m=0$ \cite{deForcrand:2014tha}.\\ 
Right: Mass dependence of the chiral and baryon
onset transition in the strong coupling limit \cite{Kim:2016izx}.}
\label{fig:efft_chiral}
\end{figure}
In this case the starting point is the lattice action with staggered fermions. After a strong coupling expansion in powers of $\beta$,
the gauge integration is done to leave a fermionic theory,
\bea
Z&=&\int D\bar{\psi}D\psi DU\;e^{-S_g[U]-S_f[U,\bar{\psi},\psi]}=\int D\bar{\psi}D\psi\; Z_f \langle e^{-S_g[U]}\rangle_{Z_f}\;, \\
Z_f&=&\int DU\;e^{-S_f[U,\bar{\psi},\psi]}\;,\quad \langle e^{-S_g[U]}\rangle_{Z_f}=1+\langle S_g[U]\rangle_{Z_f}+O(\beta^2)\;. \nn
\eea
Early mean field \cite{Kawamoto:1981hw} and Monte Carlo \cite{Karsch:1988zx} studies based on a polymer representation
have been restricted to the strong coupling limit, $\beta=0$. More recent simulations are done with a worm 
algorithm \cite{deForcrand:2009dh}.
Note that anisotropic lattices are necessary for $\beta=0$ in order to tune temperature. 

After integrating over the fermions also, one arrives at a dual formulation in terms of monomers, dimers, world lines and world sheets,
which for isotropic lattices reads
\cite{Gagliardi:2017uag,Gagliardi:2019cpa},
\beq
Z(m_q,\mu)=\sum_{\{k,n,l,n_p\}}\underbrace{\prod_{b=(x,\nu)}\frac{(N_c-k_b)!}{N_c!(k_b-|f_b|)!}}_{\tiny\mbox{meson hoppings}}
\underbrace{\prod_x\frac{N_c!}{n_x!}(2am_q)^{n_x}}_{\tiny\mbox{chiral condensate}}
\underbrace{\prod_{l_3}w(l_3,\mu)\prod_{l_f} \tilde{w}(l_f,\mu)}_{\tiny\mbox{baryon hoppings}}
\underbrace{\prod_p\frac{(\frac{\beta}{2N_c})^{n_p+\bar{n}_p}}{n_p!\bar{n}_p!}}_{\tiny\mbox{gluon propagation}}\;.\nn
\eeq
Note that this formulation in general also contains negative weights, but the resulting sign problem is mild enough to be handled
by reweighting techniques. A particular advantage of this formulation is the feasibility to simulate the chiral limit as well as
finite mass, a drawback is that gauge corrections are more difficult to include.

\fig\ref{fig:efft_chiral} (left) shows the phase diagram for the chiral
limit, both for $\beta=0$ and with leading linear gauge corrections included \cite{deForcrand:2014tha}. As expected, there always 
is a non-analytic chiral phase transition, with 
a tricritical point where the first-order transition at finite density meets the second-order line. 
In the strong coupling limit, this tricritical point  coincides with the end point of the nuclear  liquid gas transition. 
When gauge corrections are switched on,
these start splitting up, but surprisingly the first-order lines of the chiral and nuclear transitions are still indistinguishably close.
\fig\ref{fig:efft_chiral} (right) shows the strong coupling limit, but now with finite quark mass switched on. The second-order 
transition line changes to crossover, as expected. Note the decreasing $T_c(m)$ of the end point, which 
is in qualitative agreement with the finding for heavy quarks. 
The end point quickly moves to $\mu_B\gsim 3T$, which is again
consistent with all previously reported results. 
First simulations including both,
gauge corrections as well as finite mass, are also available \cite{kim}.
Note also that, in the continuum, this effective theory represents $N_f=4$ QCD 
if no rooting is applied.

\section{Conclusions}

Because direct simulations are impossible, determining the nature of the QCD transition in the chiral limit as well as 
at finite baryon density remains extraordinarily difficult. Nevertheless, systematic studies of QCD transitions in 
accessible regions of parameter space are beginning to constrain the possible phase diagram. 
In particular, the strength of the chiral transition
weakens with decreasing $N_f$ and with decreasing lattice spacing. For the chiral limit with $N_f=2$ and surprisingly also for $N_f=3$, 
this implies either a second-order transition, or a first-order transition disappearing at excessively small quark masses. The chiral transition  
also weakens, at least initially, when a real baryon chemical potential is switched on. 
For physical quark masses, there is no conclusive sign of criticality from the lattice for $\mu_B\lsim 3T$. This is consistent with
recent Dyson-Schwinger \cite{Fischer:2018sdj} and functional renormalisation group \cite{Fu:2019hdw} results, which predict a critical point 
in the range $\mu_B/T\approx 4-6$, with however still uncertain systematics in that chemical potential range.
  
\begin{figure}[t]
\vspace*{-0.8cm}
\centering
\includegraphics[height=4.5cm]{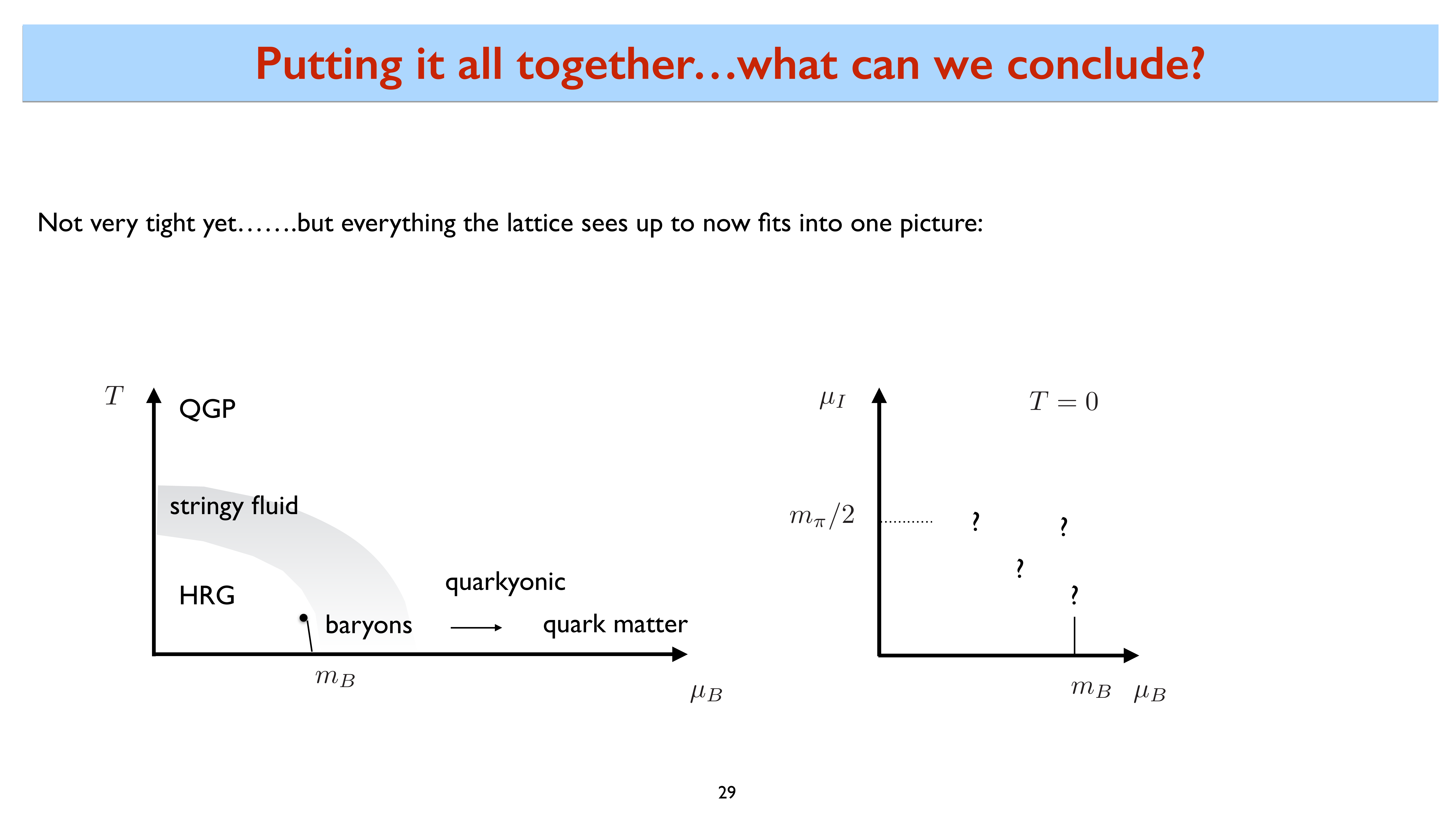}
\includegraphics[height=4.5cm]{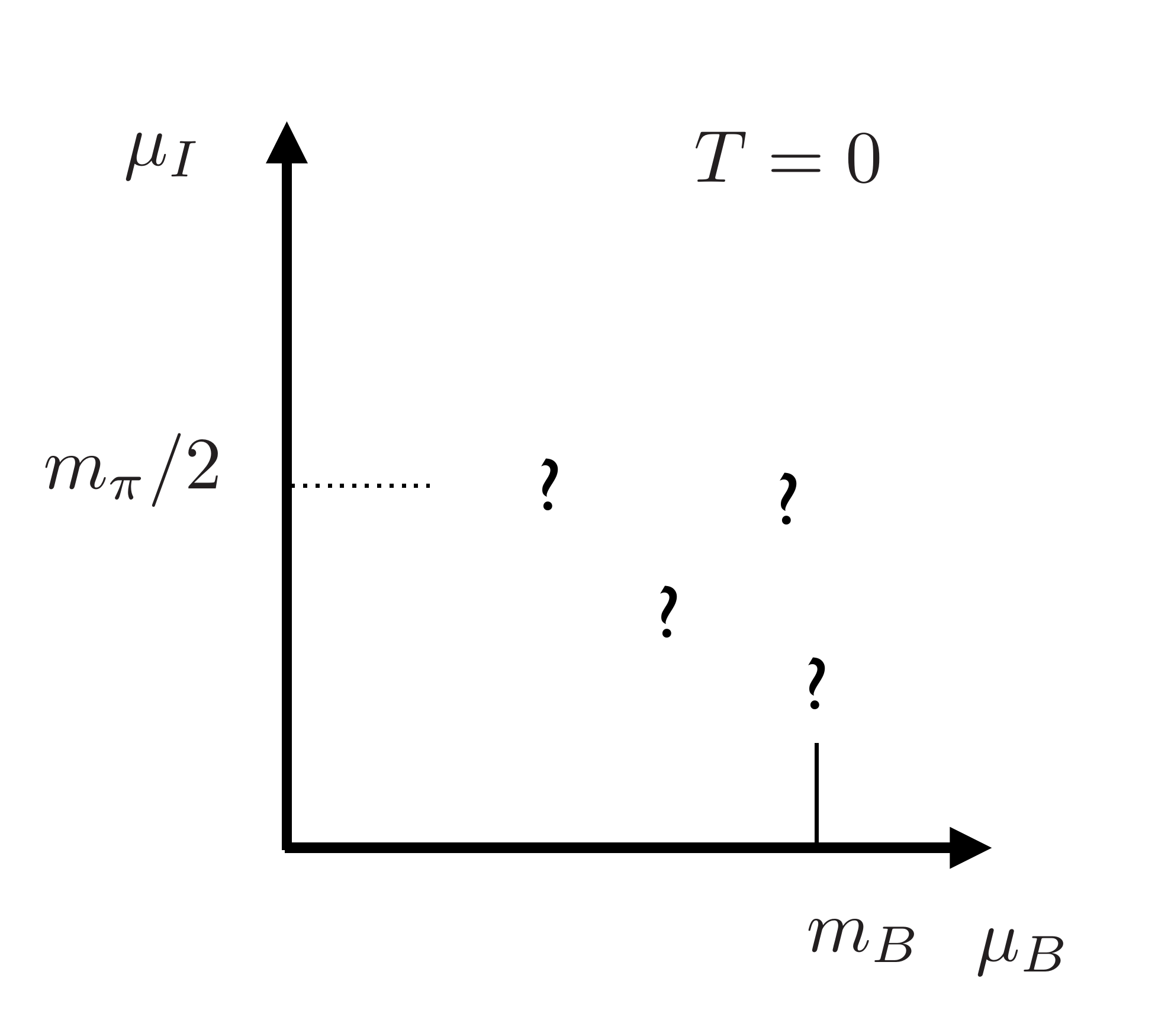}
\vspace*{-0.4cm}
\caption[]{Phase diagram with only those qualitative features, which are seen in lattice simulations so far.}
\label{fig:conc}
\end{figure}

The cold and dense regime $\mu_B/T\gg 1$, progress is being made with analytically 
derived effective lattice theories, 
which represent QCD in complementary parameter regions with either heavy quarks or at strong coupling.
They unambiguously show the 
silver blaze behaviour at $T=0$, followed by a first-order transition to baryon condensation with a critical end point,
which for physical parameter values will represent the nuclear liquid gas transition. At finite isospin chemical potential and low
temperatures, a second-order transition to a pion condensed phase is seen for physical quark masses and in the continuum. 
These features are indicated in the phase diagram \fig\ref{fig:conc}. 

Other developments concern the physical degrees of freedom near the thermal transition. An emergent chiral spin symmetry
in the temperature range from the crossover to $\sim 1$ GeV suggests light quarks still bound by colour-electric strings, 
and the symmetry gets amplified by baryon chemical potential. At low temperatures, there is the onset transition to baryon matter, which
smoothly turns into a quarkyonic regime (defined by $p\sim N_c$) that, at least in principle, allows a continuous interpolation from
baryon matter to quark matter. The band in the phase diagram  \fig\ref{fig:conc} thus indicates a region, 
where the dynamics changes very gradually
and the degrees of freedom still resemble the hadronic ones. In conclusion, the lattice is beginning to see some 
structure in the QCD phase diagram. Intriguingly, a non-analytic chiral phase transition is neither required nor 
ruled out by lattice data at this stage, and remains an exciting subject of research.   

\newpage
\acknowledgments 
The author is grateful to B.~Brandt for helpful discussions and acknowledges support by the Deutsche Forschungsgemeinschaft (DFG) 
through the grant CRC-TR 211 ``Strong-interaction matter
under extreme conditions'', as well as by the Helmholtz International Center for FAIR within the LOEWE program of the State of Hesse.


\end{document}